\documentclass{aastex62}

\newcommand{\upp}

\begin{document}
\title{The Second Plateau in X-ray Afterglow Providing Additional Evidence for Rapidly Spinning Magnetars as the GRB Central Engine\\Submitted to ApJ 2020 March 10; Accepted 2020 May 1}

\correspondingauthor{He Gao}
\email{gaohe@bnu.edu.cn}
\correspondingauthor{Liangduan Liu}
\email{liuliangduan@bnu.edu.cn}

\author[0000-0002-0786-7307]{Litao Zhao}
\affil{Department of Astronomy ,
Beijing Normal University, Beijing, China;}

\author[0000-0002-8708-0597]{Liangduan Liu}
\affiliation{Department of Astronomy ,
Beijing Normal University, Beijing, China;}

\author[0000-0002-3100-6558]{He Gao}
\affiliation{Department of Astronomy ,
Beijing Normal University, Beijing, China;}

\author{Lin Lan}
\affiliation{Department of Astronomy ,
Beijing Normal University, Beijing, China;}

\author[0000-0003-3440-1526]{WeiHua Lei}
\affiliation{School of Physics, Huazhong University of Science and Technology, Wuhan, China}

\author{Wei Xie}
\affiliation{Guizhou Provincial Key Laboratory of Radio Astronomy and Data Processing, Guizhou Normal University, Guiyang, 550001, China.}

\begin{abstract}

Evidence for the central engine of gamma-ray bursts (GRBs) has been collected in the Neil Gehrels \emph{Swift} data. For instance, some GRBs show an internal X-ray plateau followed by very steep decay, which is difficult to be interpreted within the framework of a black hole (BH) central engine, but are consistent within a rapidly spinning magnetar engine picture. The very steep decay at the end of the plateau suggests a sudden cessation of the central engine, which is explained as the collapse of a supra-massive magnetar into a black hole when it spins down. Here we propose that some additional evidence, such as a second X-ray plateau feature would show up, if the fall-back accretion could activate the newborn BH and sufficient energy could be transferred from the newborn BH to the GRB blast wave. With a systematic data analysis for all long GRBs, we find three candidates in \emph{Swift} sample, i.e., GRBs 070802, 090111, and 120213A, whose X-ray afterglow light curves contain two plateaus, with the first one being an internal plateau. We find that in a fairly loose and reasonable parameter space, the second X-ray plateau data for all 3 GRBs could be well interpreted with our proposed model. Future observations are likely to discover more similar events, which could offer more information of the properties of the magnetar as well as the newborn BH.
\end{abstract}


\section{Introduction} \label{sec:intro}

Gamma-ray bursts (GRBs) have been extensively explored since their discovery more than 50 years ago, but the nature of the GRBs central engine remains a mystery. In the literature, two main kinds of central engine have been well discussed: hyper-accreting black holes (BH) or rapidly spinning magnetars \cite[][for a review]{zhangbook}. It has long been proposed that some interesting signatures showing in some GRB's X-ray afterglow could help us to determine their central engines \citep{dailu98,rees98,zhangmeszaros01,zhang06,nousek06}. For instance, systematic analysis for the Swift GRB X-ray afterglow shows that bursts with X-ray plateau features likely have rapidly spinning magnetars as their central engines \citep{liang07,zhao19,tang19}. In particular, when the X-ray plateau is followed by a steep decay with temporal decay index $\alpha\gtrsim3$, hereafter called ``internal X-ray plateaus", the sharp decay at the end of the plateau is difficult to be interpreted within the framework of a BH central engine, but are consistent within a magnetar engine picture, where the abrupt decay is understood as the collapse of a supra-massive magnetar into a BH after the magnetar spins down \citep{troja07,lyons10,rowlinson10,rowlinson13,luzhang14,lu15,gao16a,depasquale16,ZhangQ16}. 

Recently, \cite{chen17} proposed that if the sudden drop after the internal plateau indeed indicates the collapse of a supra-massive NS into a BH, signatures from this newborn BH should be expected. For instance, for long GRBs if a fraction of the envelope material fall back and activate the accretion onto the newborn BH \citep{kumar08a,kumar08b,wu13,gao16b}, the hyper-accreting BH system can launch a relativistic jet via Blandford-Znajek (BZ) mechanism \citep{BZ77}. Supposing a fraction $\xi$ of the jet energy would undergo internal dissipation, some detectable signals, such as a X-ray bump following the internal plateau is expected \citep{chen17}. Searching through \emph{Swift} -XRT data archive, the authors found a particular case, GRB 070110, which showed a small X-ray bump following its internal plateau, and successfully interpreted its multi-band data with their model. But it is worth noting that for GRB 070110, \cite{troja07} found the optical data indicate a decaying light curve feature sitting underneath the X-ray plateau without showing the rapid drop for the ``internal plateau", which means unlike the X-ray band, the optical emission trend could be hardly disturbed by the late central engine activity.

Note that most internal plateaus are not found to be followed by the X-ray bump, it may be because that the late time fallback processes are intrinsically weak or the energy fraction for the internal dissipation ($\xi$) is relatively small. For the latter situation, a food fraction $1-\xi$ of the BZ jet energy would continuously inject into the GRB blast wave, which may generate a second plateau in X-ray afterglow, if the injected energy is comparable or even larger than the blast wave kinetic energy. We thus propose that GRBs with two X-ray plateaus (the first one is internal plateau) may provide further support to the magnetar central engine model. 

In this work, we first study how the energy injection from the fallback accretion onto the newborn BH would alter the GRB afterglow emission and analyze the influence of model parameters on the theoretical lightcurves. In section 3, we systematically search for GRBs with two X-ray plateaus from the Swift-XRT sample. Considering that long GRBs, which are likely related to the core-collapse of massive stars \citep{woosley93,macfadyenwoosley99}, are more easier to have enough envelop material to provide late time fall-back accretion than short GRBs, that have been proposed to originate from the merger of two neutron stars \citep{eichler89,narayan92} or the merger of a neutron star and a black hole \citep{paczynski91}, here we focus on long GRB samples\footnote{Some previous works found some short GRBs with two X-ray plateaus and explained the data with their own model. For instance, \cite{hou18} proposed that two plateaus showing in the X-ray light curve of GRB 170714A could be interpreted as the solidification and collapsing process of a hyper-massive quark star. \cite{zhangQ18} proposed that a second plateau showing in GRB 160821B could be interpreted by fall-back accretion process from a magnetically-arrested disk}. We find three candidates, GRB 070802, GRB 090111 and GRB 120213A, whose second plateaus could be well fitted by the model. The conclusion and implications of our results are discussed in Section 4. Throughout the paper, the convention $Q=10^nQ_n$ is adopted in c.g.s. units.

\section{Model Description} \label{sec:model}

Rapidly spinning magnetars have long been proposed as one of the candidate of GRB central engines. In this scenario, a collimated jet could be launched by invoking (1) hyper-accretion onto the NS \citep{zhangdai09,zhangdai10,bernardini13}; (2) magnetic bubbles from a differentially millisecond proto-NS \citep{dai06}; (3) or from a protomagnetar wind \citep{metzger11}. The internal dissipation of the jet could power the prompt gamma ray emission of a GRB and the interaction between the jet and the ambient medium could produce a strong external shock that gives rise to bright broadband afterglow emission \cite[][for a review]{gao13}. After launching the jet, the magnetar would also eject a near-isotropic Poynting-flux-dominated outflow, the internal dissipation of which could power a bright X-ray emission, whose temporal profile would follow the spin-down profile of the magnetar, i.e., $t^{0}$ at early stage and decay as $t^{-2}$ after the magnetar spinning down \citep{zhangmeszaros01}. If this emission component is brighter than the external shock afterglow emission, an X-ray plateau would show up. Sometimes the central engine magnetar might be a supra-massive NS, which would collapse to a BH when a good fraction of its rotational energy is lost and the centrifugal support can no longer support gravity. In this case, the X-ray plateau emission would suddenly stop and follows by a sharp decay at the end of the plateau, due to the abrupt cessation of the magnetar’s central engine. This explains the internal X-ray plateau discovered in both long and short GRBs \citep{troja07,lyons10,rowlinson10,rowlinson13,luzhang14,lu15}. After the collapse of the magnetar, if a fraction of the envelope material (especially for long GRBs) fall back and activate the accretion onto the newborn BH, the rotational energy of the BH could be extracted via BZ mechanism, and a good fraction of the energy would eventually inject into the afterglow blast wave. If the injected energy is comparable or even larger than the blast wave kinetic energy, the broadband afterglow lightcurve could be significantly altered, for instance, a second plateau in the X-ray afterglow may emerge. Figure 1 presents a physical picture for several emission components at different temporal stages. Here we focus on study how the energy injection from the fallback accretion onto the newborn BH would alter the GRB afterglow emission and analyze the influence of model parameters on the theoretical lightcurves.

\begin{figure}[!htb]
\centering
\includegraphics[width=12cm,height=9cm]{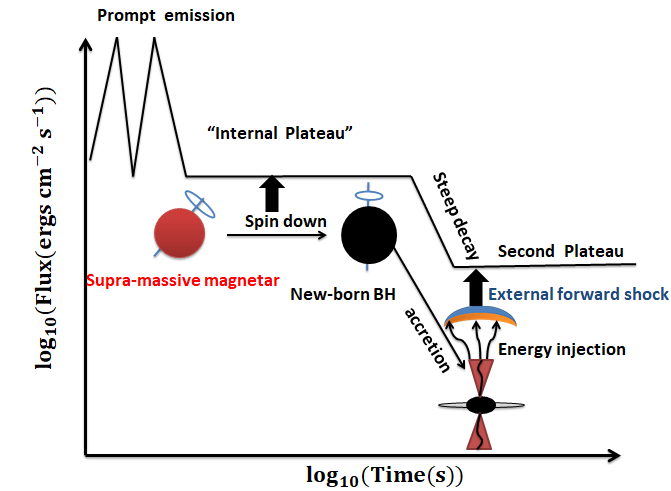}
\caption{Model illustration: the ``internal plateau'' is powered by the spin-down power from a supra-massive magnetar and the steep decay marks the collapse of the magnetar into a BH when it spins down. In addition, the second plateau is caused by energy transfer from new born BH through BZ mechanism to GRB blast wave.}
\label{fig-2}
\end{figure}

\subsection{The Fallback Accretion onto the Newborn BH}\label{fallback}

Assuming the fallback accretion could trigger the energy extraction from the newborn BH via BZ mechanism, in this case, the BZ power from a BH with mass $M_{\bullet}$ and angular momentum $J_{\bullet}$ could be estimated as \citep{Lee00,Li00,wang02,lei05,lei13,lei17,mckinney05,Leizhang13,chen17,liu17,Lloyd18}

\begin{eqnarray}
L_{\rm{BZ}}=1.7 \times 10^{50} a^{2} m_{\bullet}^{2} B_{\bullet, 15}^{2} F(a) \,\rm{erg \ s}^{-1},
\label{eq-1}
\end{eqnarray}
where $ m_{\bullet}=M_{\bullet}/M_{\odot}$ is the dimensionless BH mass and $a=J_{\bullet}c /(G M_{\bullet}^{2})$ is the dimensionless spin parameter of the BH. Here, $F(a) =[(1+q^{2}) / q^{2}][(q+1 / q) \arctan q-1]$, $q=a /(1+\sqrt{1-a^{2}})$, and $B_{\bullet,15}$  is the magnetic-field strength threading the BH horizon in units of $10^{15}$ G. The evolution of the BZ jet power depends on  $m_{\bullet}$, $a$, and $B_{\bullet}$.

The evolution of the BH spin and mass governed the competition between spin up by accretion and spin down by the BZ mechanism. The evolution equations of the BH mass $M_{\bullet}$ and the BH spin $a$  are given by \cite{wang02}
\begin{eqnarray}
\frac{d M_{\bullet} c^{2}}{d t}=\dot{M} c^{2} E_{\rm{ms}}-L_{\rm{BZ}},
\end{eqnarray}
and 
\begin{eqnarray}
\frac{d a }{d t}=\frac{(\dot{M} L_{\rm{ms}}-T_{\rm{BZ}}) c}{G M_{\bullet}^{2}}-\frac{2 a(\dot{M} c^{2} E_{\rm{ms}}-L_{\rm{BZ}})}{M_{\bullet}c^{2}},
\end{eqnarray}
where $\dot M$ is the BH accretion rate,  and $T_{\rm{BZ}}$ is BZ magnetic torque \citep{Li00,Leizhang13,lei17}
\begin{eqnarray}
T_{\rm{BZ}}=3.36 \times 10^{45} a^{2} q^{-1} m_{\bullet}^{3} B_{\bullet, 15}^{2} F(a) \rm{g} \,\rm{cm}^{2}\,\rm{s}^{-2},
\end{eqnarray}
where $E_{\rm{ms}}$ and $L_{\rm{ms}}$  are  the  specific energy and the specific angular momentum at innermost radius $r_{\rm{ms}}$  of the disk \citep{novikov73}:
\begin{eqnarray}
E_{\rm{ms}}=(4 \sqrt{R_{\rm{ms}}}-3 a) /(\sqrt{3} R_{\rm{ms}}),\\
L_{\rm{ms}}=(GM_{\bullet}/c)(2(3 \sqrt{R_{\rm{ms}}}-2 a)) /(\sqrt{3}\sqrt{R_{\rm{ms}}}).
\end{eqnarray}
Here, $R_{\rm{ms}}=r_{\rm{ms}} / r_{\rm{g}}$ and $ r_{\rm{g}}= G M_{\bullet} / c^{2}$. The  innermost stable radius of disk is \citep{bardeen72}
\begin{eqnarray}
r_{\rm{ms}}=r_{\rm{g}}\left[3+Z_{2}-\left[(3-Z_{1})(3+Z_{1}+2 Z_{2})\right]^{1 / 2}\right],
\end{eqnarray}
where $Z_{\rm{l}} \equiv 1+(1-a^{2})^{1 / 3}\left[(1+a)^{1 / 3}+(1-a)^{1 / 3}\right] $ and $Z_{2} \equiv(3 a^{2}+Z_{1}^{2})^{1 / 2}$ for $0 \leqslant a \leqslant 1$.

The strength of the magnetic field is the major uncertainty in estimating the BZ jet power. By assuming  the ram pressure of the innermost part of the disk balances the magnetic pressure on the BH horizon (\cite{modersiki97}), one can estimate the magnetic field strength $B_\bullet$,
\begin{eqnarray}
\frac{B_{\bullet}^{2}}{8 \pi}\approx \frac{\dot{M} c}{4 \pi r_{\bullet}^{2}}.
\end{eqnarray}
where $r_{\bullet}$ is the radius of the BH horizon.

With this assumption, the BZ jet power could be written as a function mass accretion rate and BH spin, i.e. \citep{wu13,chen17}
\begin{eqnarray}
L_{\rm{BZ}}=9.3 \times 10^{53} \frac{a^{2} \dot{m}F(a)}{(1+\sqrt{1-a^{2}})^{2}} \rm{erg \ s}^{-1},
\label{eq-2}
\end{eqnarray}
where $ \dot{m}=\dot{M}/(M_{\odot} \,\rm{s}^{-1})$ is the  dimensionless BH accretion rate. The accretion rate of BH can be estimated by adopting a simple model described in \cite{kumar08a}
\begin{equation}\label{Eq:dotM}
\dot{M} \simeq \frac{M_{\rm d}}{\tau_{\rm{vis}}},
\end{equation}
where the viscous timescale $\tau_{\rm{vis}} \sim 1/\alpha \Omega_{K}$, here  $\alpha$  is the standard dimensionless viscosity parameter, $ \Omega_{K}$ is the Kepler angular velocity of accretion disk.

The mass of the disk $M_{\rm d}$  evolves with time, it increases as a result of fall-back from the envelope and decrease as a result of accretion. Thus, one has \citep{kumar08a,lei17}
\begin{equation}\label{Eq:M_d}
\dot{M}_{\rm d} = \dot{M}_{\rm{fb}} - \dot{M}.
\end{equation}

Combining Eqs  (\ref{Eq:dotM}) and (\ref{Eq:M_d}), one can obtain the accretion rate onto the BH  \citep{kumar08a,lei17},
\begin{eqnarray}\label{Eq:tau_vis}
\dot{M}=\frac{1}{\tau_{\rm{vis}}} e^{-t / \tau_{\rm{vis}}} \int_{t_{0}}^{t} e^{t^{\prime} / \tau_{\rm{vis}}} \dot{M}_{\mathrm{fb}} d t^{\prime}.
\end{eqnarray}

The evolution of the fall-back accretion rate is described with a broken-power-law  as \citep{chevalier89,macfadyen01,zhangwei08,Dailiu12}
\begin{eqnarray}\label{Eq:fallback}
\dot{M}_{\rm{fb}}=\dot{M}_{\rm{p}}\left[\frac{1}{2} \left(\frac{t-t_{0}}{t_{\rm{p}}-t_{0}} \right)^{-1 / 2}+\frac{1}{2} \left(\frac{t-t_{0}}{t_{\rm{p}}-t_{0}} \right)^{5 / 3}\right]^{-1},
\end{eqnarray}
where $t_{0}$ is the start time of the fall-back accretion in the local frame, $t_{\rm p}$ is peak time of fallback and and $\dot{M}_{\rm p}$ is the peak fall-back rate.  

For the rapid accretion case, $\tau_{\rm{vis}} \ll t$, the BH accretion rate would follow the fall-back rate, i.e., $\dot{M} = \dot{M}_{\rm fb}$. For a large value of the viscosity timescale $\tau_{\rm{vis}}$, the BH accretion rate would be flat until $t>\tau_{\rm{vis}}$, and then starts to decline with time, see Figure 7 in \cite{lei17}.

\begin{figure*}
	\begin{center}
		\setlength{\abovecaptionskip}{0.cm}
		\setlength{\belowcaptionskip}{-0.cm}
		\hspace{0cm}
		\includegraphics[width=5.5cm,height=5cm]{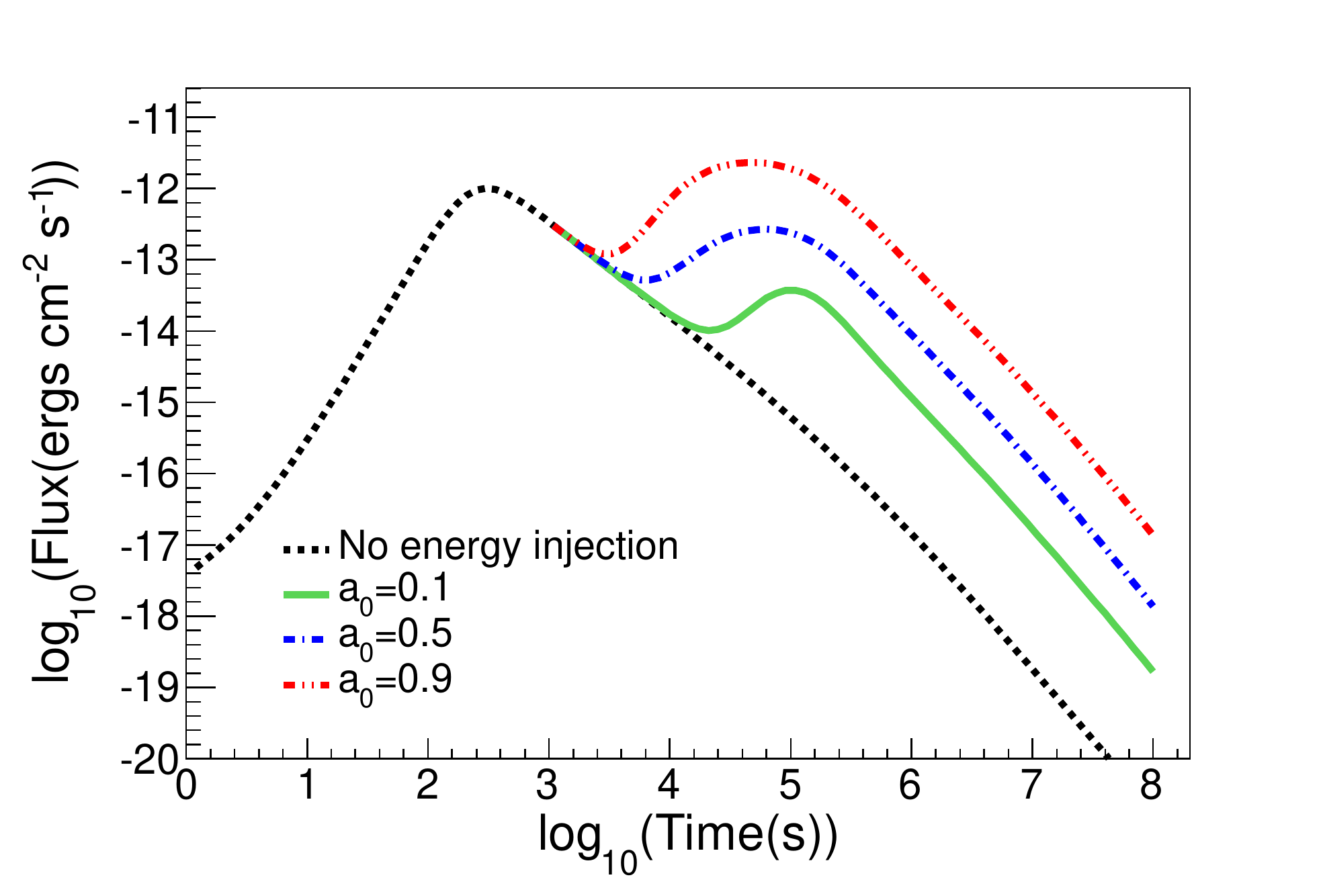}
		\includegraphics[width=5.5cm,height=5cm]{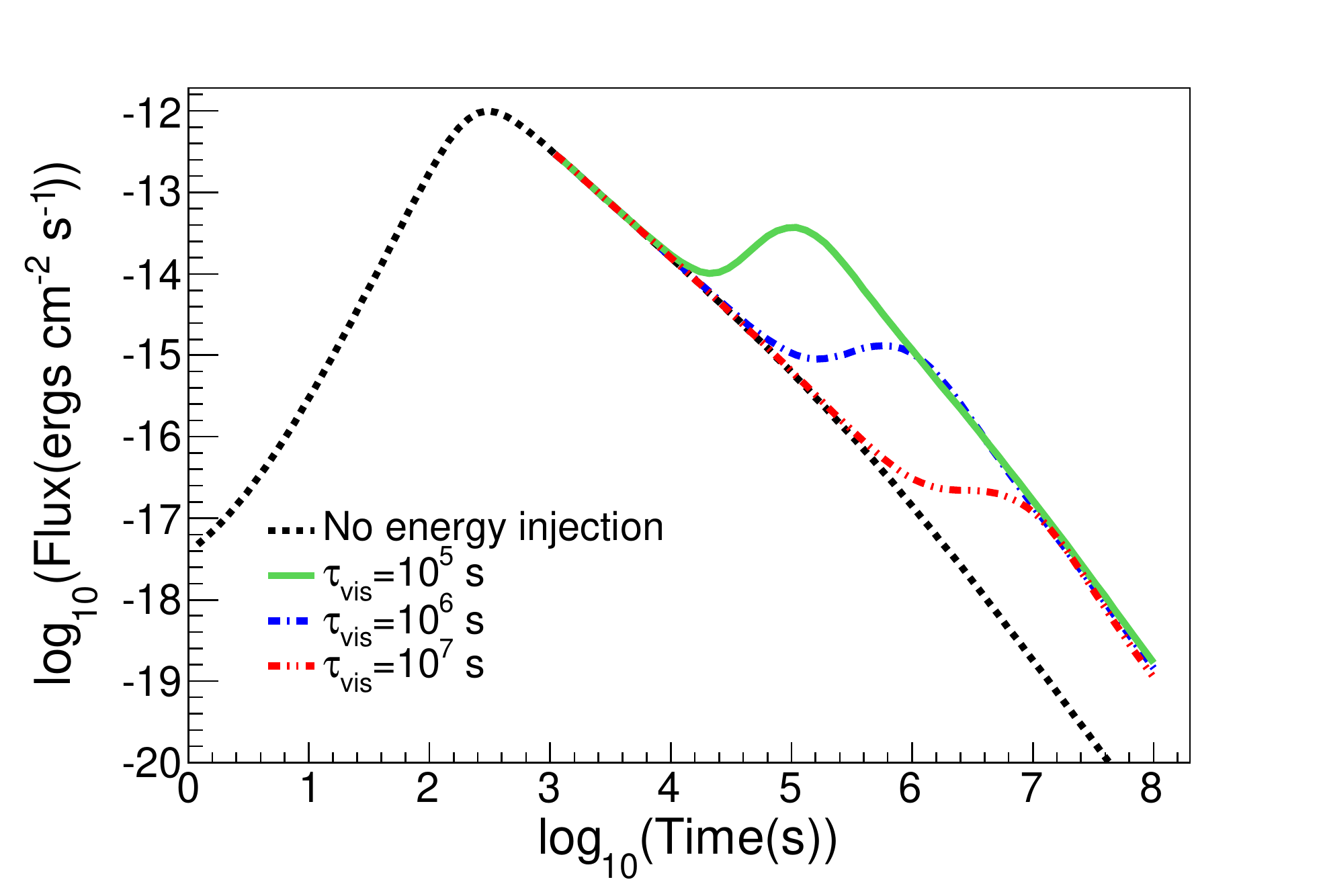}
		\includegraphics[width=5.5cm,height=5cm]{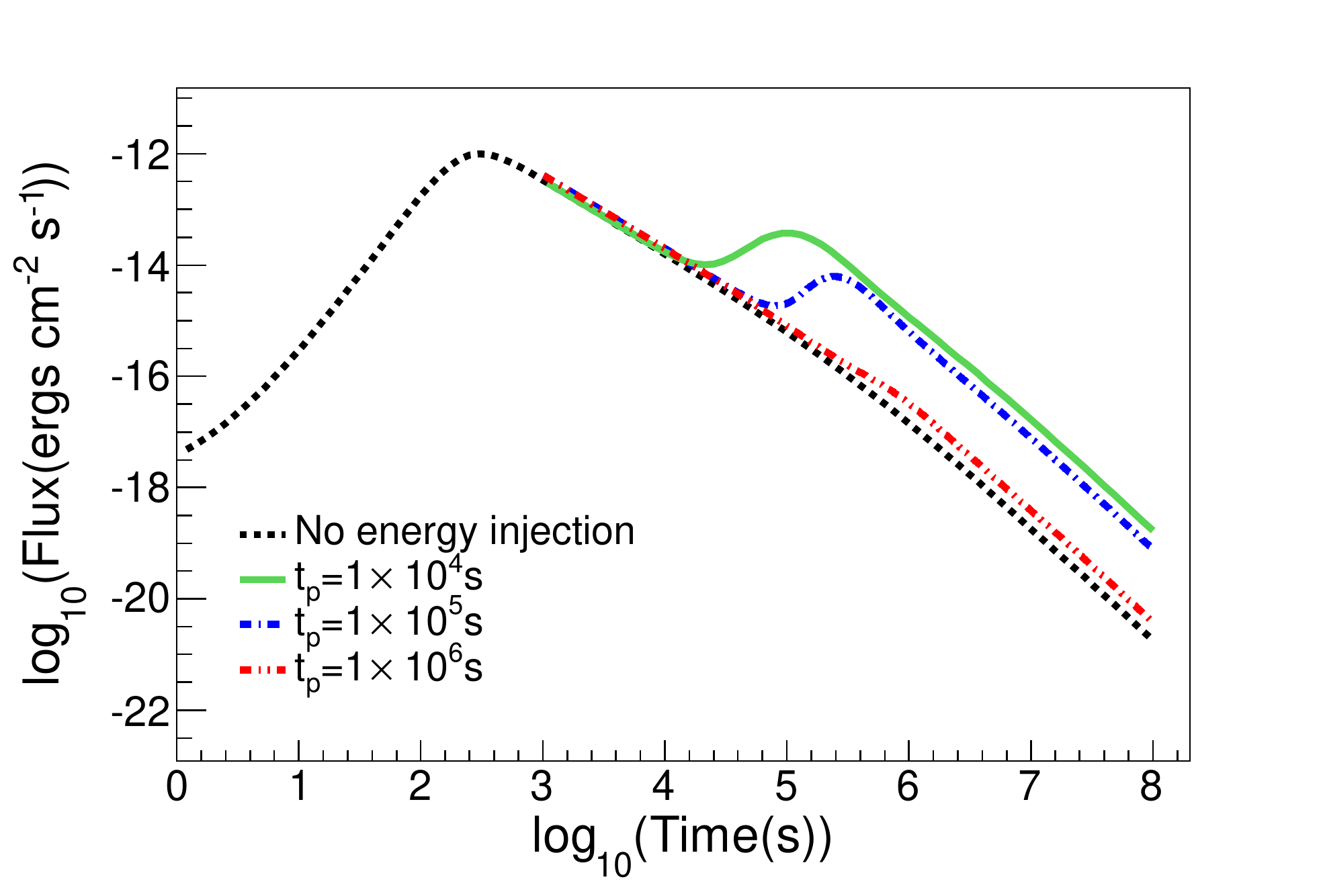}
		\includegraphics[width=5.5cm,height=5cm]{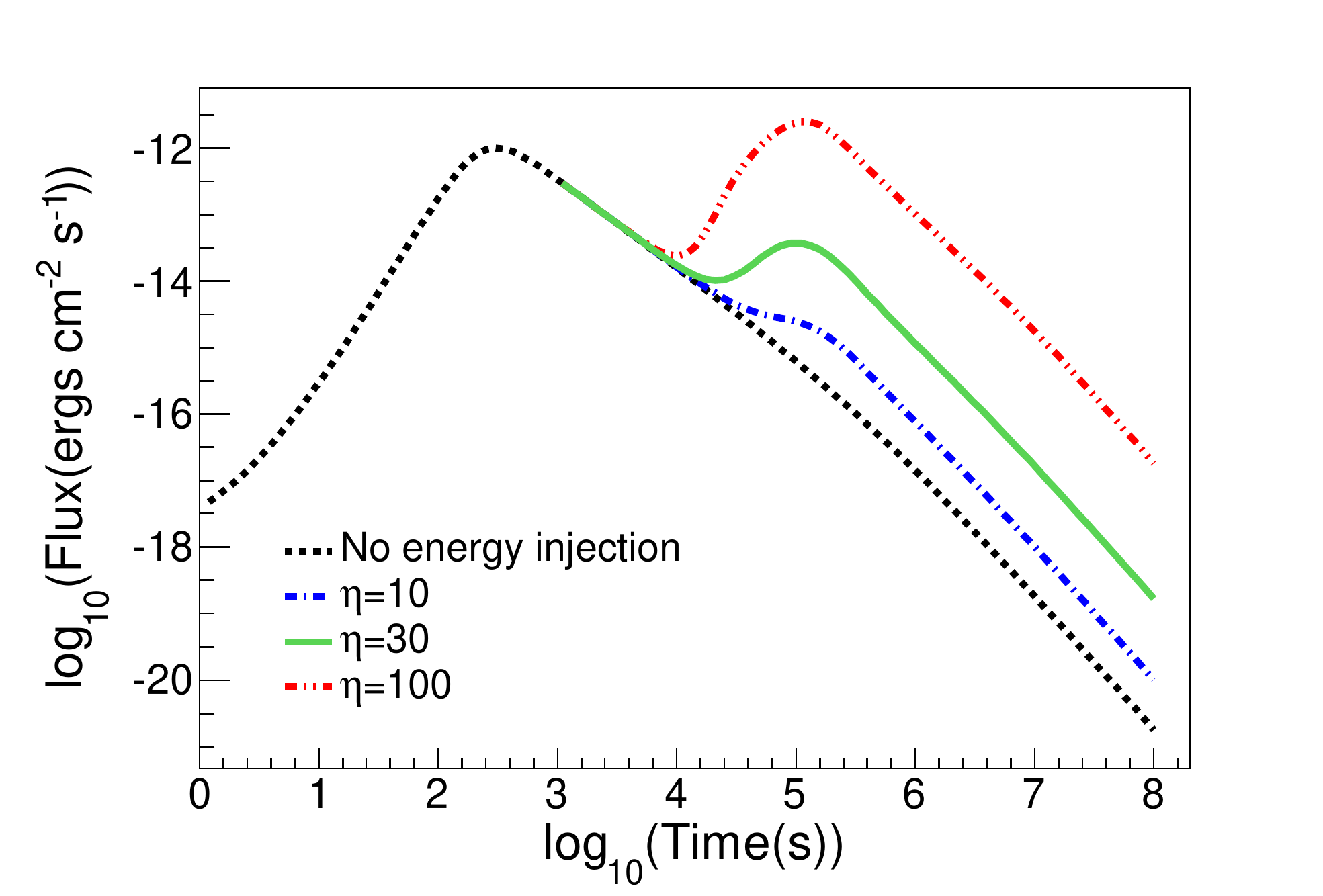}
		\includegraphics[width=5.5cm,height=5cm]{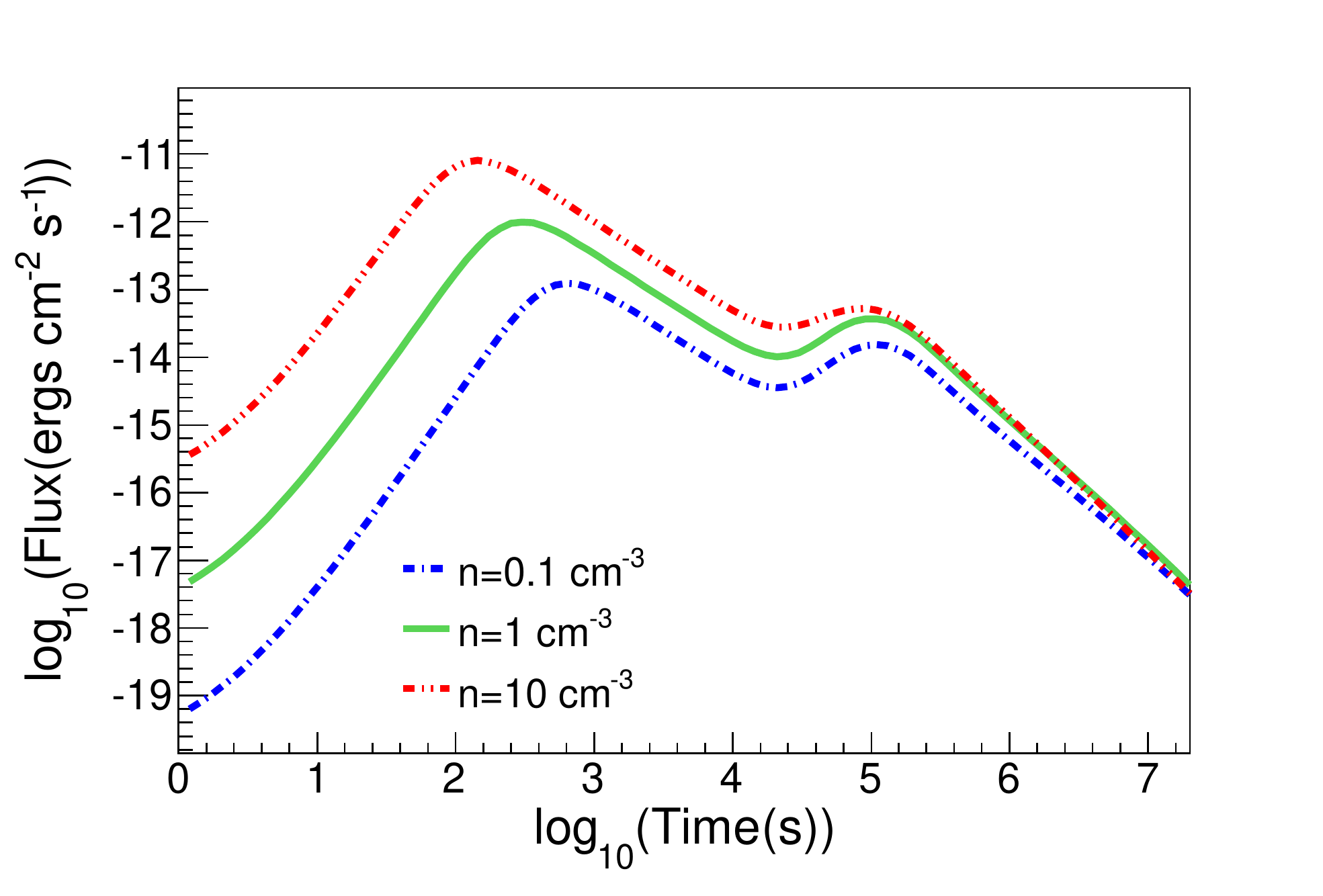}
		\includegraphics[width=5.5cm,height=5cm]{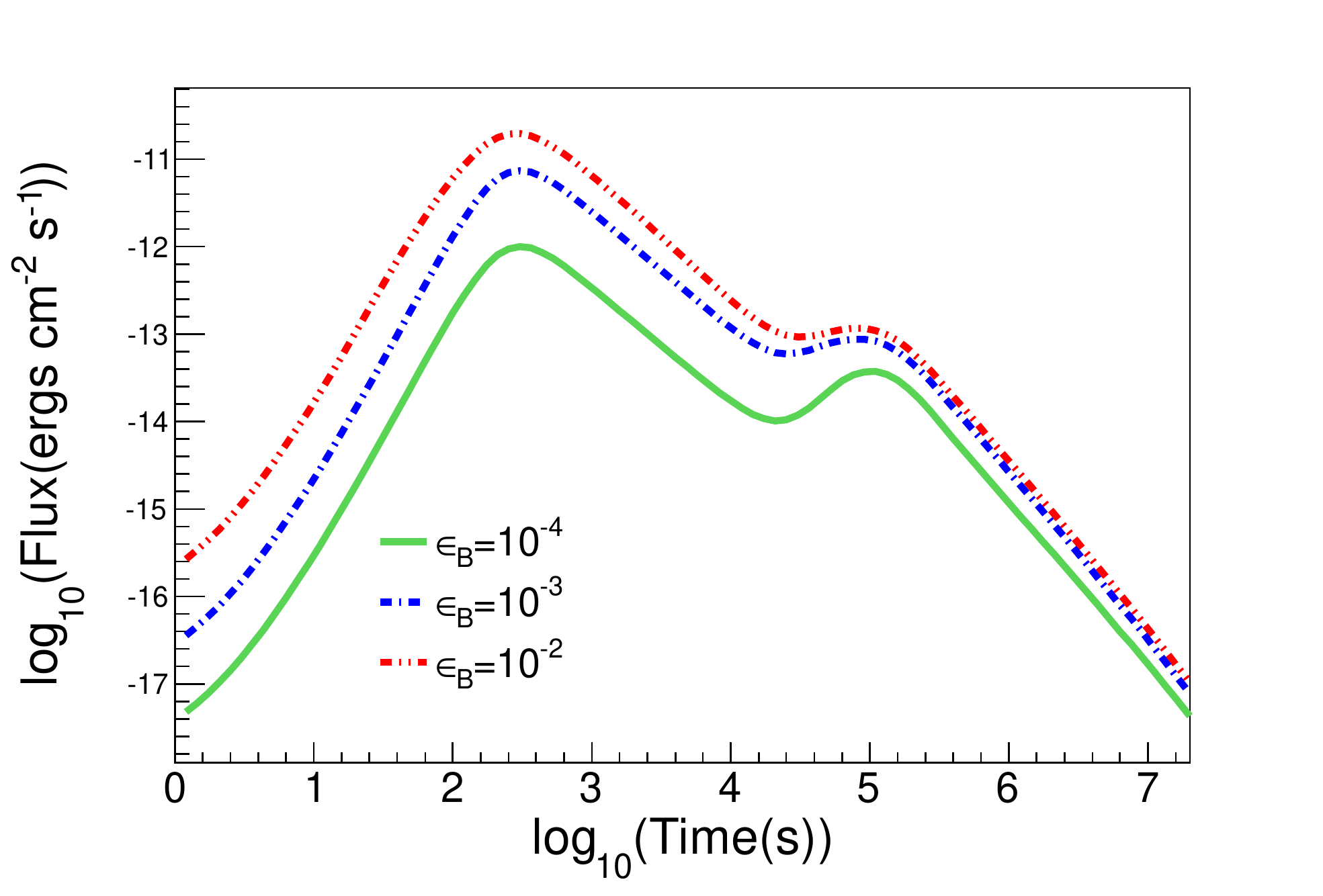}
		\includegraphics[width=5.5cm,height=5cm]{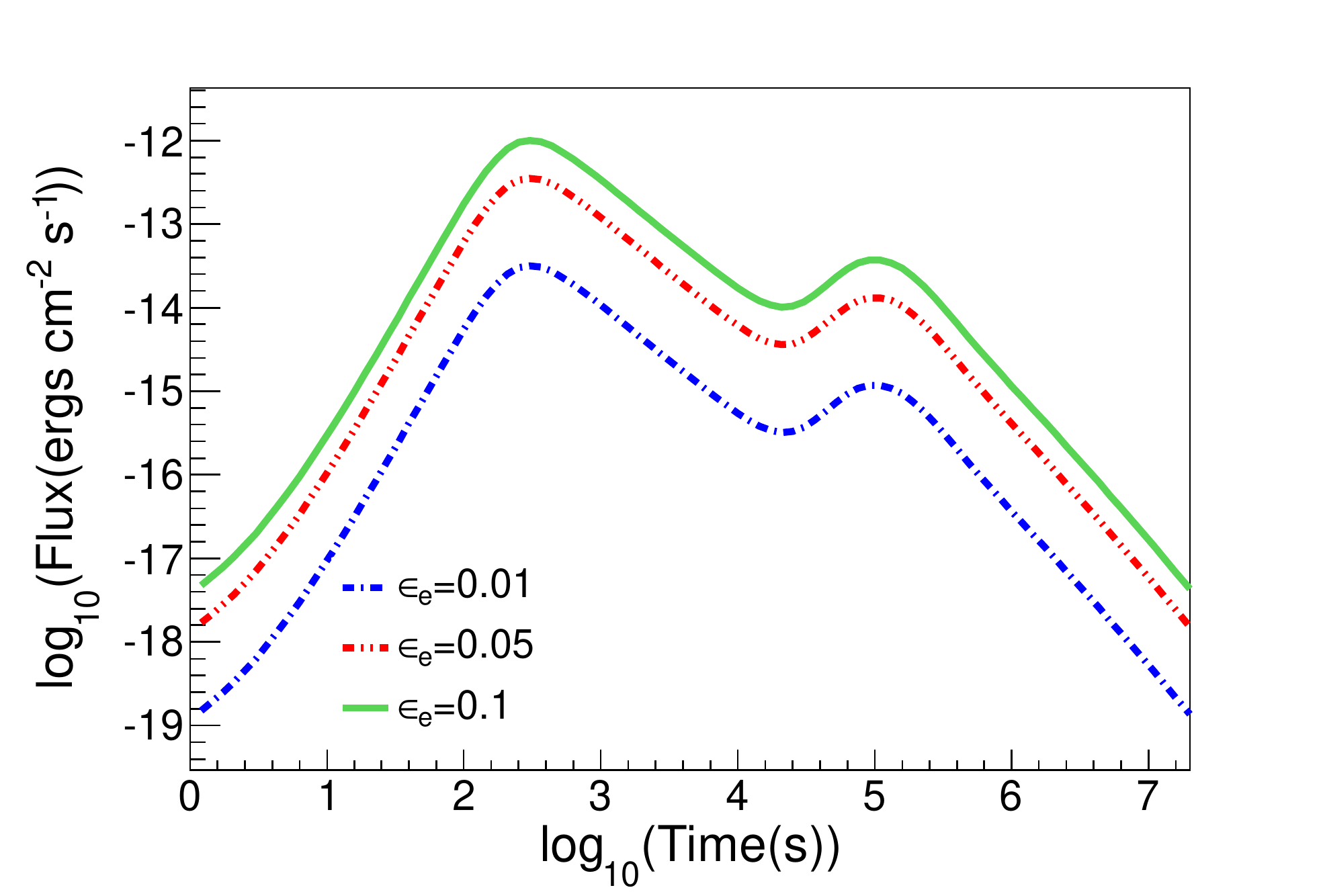}
		\includegraphics[width=5.5cm,height=5cm]{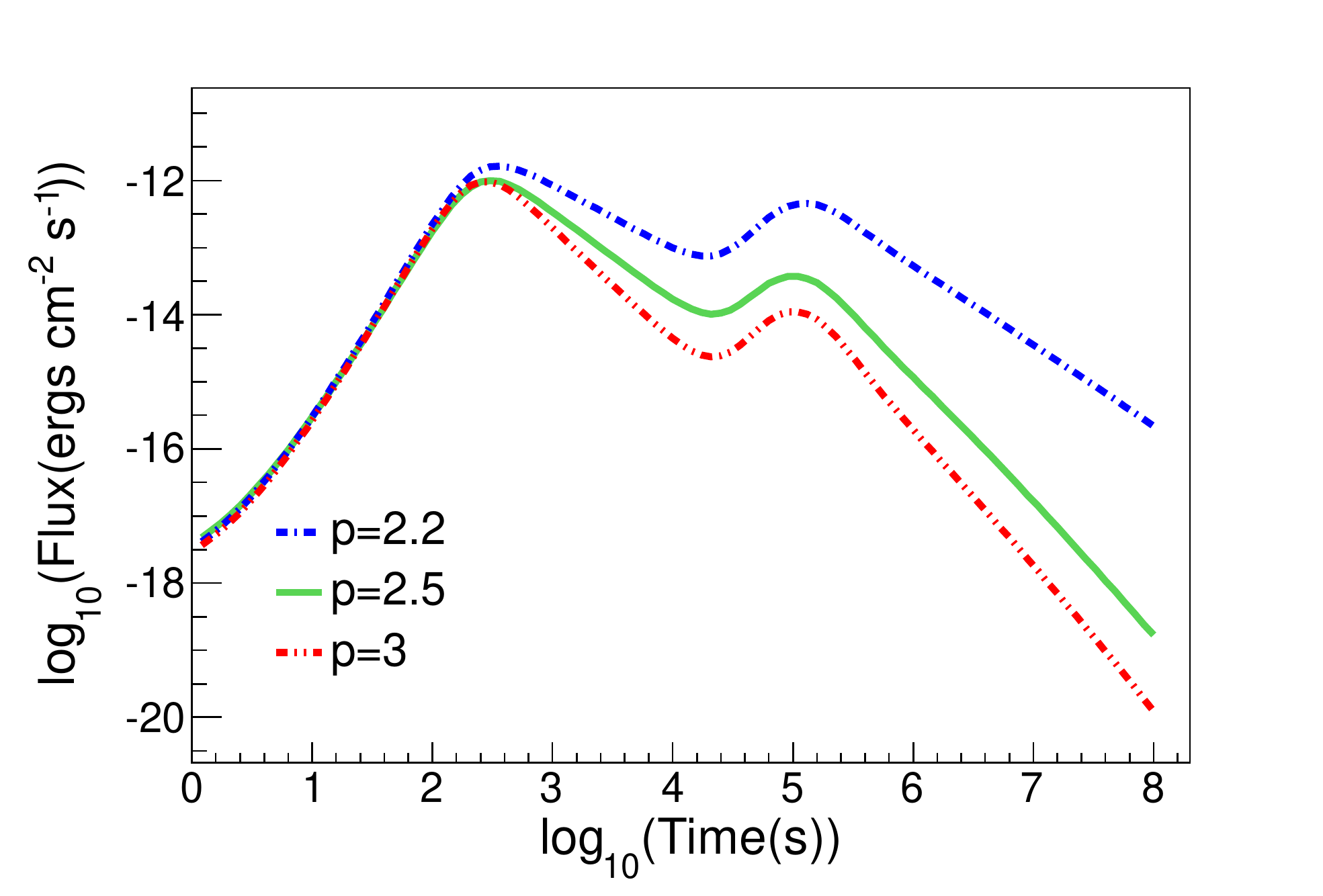}
		\includegraphics[width=5.5cm,height=5cm]{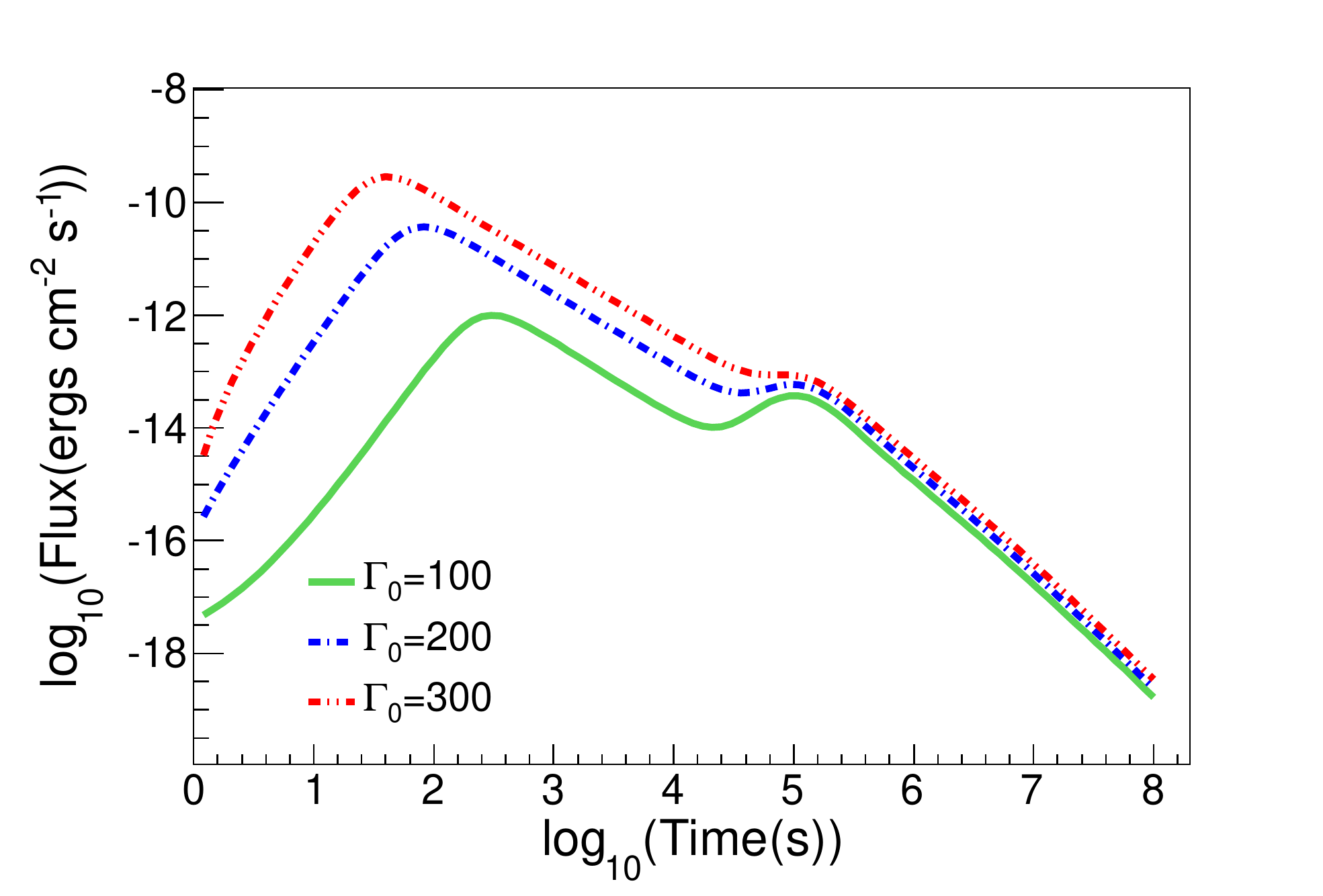}
		\caption{Influence of the model parameters on the X-ray light curve. Except as noted in each subfigure, in the calculation, we take a set fiducial values for the model parameters: $E_{0}=10^{52}\ \rm{ergs}$,  $\Gamma_{0}=100$, $n =1\ \rm{cm}^{-3}$, $\theta_{0}=0.1$,$\epsilon_{e}=0.1$, $\epsilon_{B}=10^{-4}$, $p=2.5$,  $a_0=0.1$, $\tau_{\rm{vis}}=10^{5}$ s, $t_p=10^{4}$ s and $\eta=30$.}
		\label{fig-3}
	\end{center}
\end{figure*}

\subsection{Energy injection into the GRB afterglow blast wave}

The energy flow from the BZ process would continuously inject into the external shock, and cause a significant raise to the Lorentz factor of the blast wave $\Gamma$, which may produce the second plateau following the steep decay.  
\cite{huang00} proposed a generic dynamical model to describe the dynamical evolution of GRB outflow, which has been widely applied for modeling the afterglow light curve. Based on their model and taking the energy injection into account, the evolution equation of the outflow's bulk Lorentz factor can be written as \citep{Liu2014}
\begin{eqnarray}
\frac{d \Gamma}{d M_{\rm{sw}}}=-\frac{1}{M_{\rm{ej}}+2 \Gamma M_{\rm{sw}}}\left[\Gamma^{2}-1-\frac{L_{\rm{BZ}}}{c^{2}} \frac{dt}{d M_{\rm{sw}}   }\right],
\end{eqnarray}
where $M_{\rm sw}$ is the swept-up mass by shock, $M_{\rm ej}$ is the initial mass of the GRB outflow. The blast wave energy continuously increase with time, due to the continuously injected energy from the BZ process into the blast wave.

The initial kinetic energy of GRB outflow can be estimated as $E_0 = \Gamma_0 M_{\rm ej} c^2$, and the injected energy from the BZ process can be calculated as\footnote{Since we focus on the case where the BZ jet undergoes weak internal dissipation, here we simply set $\xi=0$.} $E_{\rm inj}=\int (1-\xi) L_{\rm BZ}~dt$. The total energy in the blast wave could be expressed as 
\begin{eqnarray}
E_{\rm tot} =E_0+E_{\rm inj}.
\end{eqnarray}
We introduce a parameter $\eta \equiv E_{\rm inj}/E_0$ to denote the ratio between the injected energy and the initial energy in blast wave. For $\eta>1$, the injected energy is dominated in the total energy of the blast wave, otherwise the initial kinetic energy of GRB outflow  is dominated.

In order to obtain the shock dynamical evolution, three additional differential equations are required \citep{huang00}. The evolution of the radius of shock $R$,  the swept-up mass $M_{\rm{sw}}$, and the opening angle of the jet $\theta$  are described by \cite{huang00}
\begin{eqnarray}
\frac{d R}{d t}=\sqrt{\frac{\Gamma^{2}-1}  {\Gamma}} c \Gamma(\Gamma+\sqrt{\Gamma^{2}-1}),\\
\frac{d M_{\rm{sw}}}{d R}=2 \pi R^{2}(1-\cos \theta) n m_{p},\\
\frac{d \theta}{d t}=\frac{c_{s}(\Gamma+\sqrt{\Gamma^{2}-1})}{R},
\end{eqnarray}
where $n$ is the number density of the unshocked ISM and $c_{s}$ is the sound speed,
\begin{eqnarray}
c_{s}^{2}=\hat{\gamma}(\hat{\gamma}-1)(\Gamma-1) \frac{1}{1+\hat{\gamma}(\Gamma-1)} c^{2},
\end{eqnarray}
where the adiabatic index $\hat{\gamma}=(4\Gamma+1)/(3\Gamma)$.

For an electron with an energy $\gamma_e m_e c^2$ in the co-moving frame of the shock, the observed frequency from sychrotron emission is \citep{rybickilightman79}
\begin{eqnarray} \label{Eq:fre}
\nu(\gamma_{e})=\frac{3}{4\pi}\Gamma \gamma_{e}^{2} \frac{q_{e} B'}{ m_{e} c},
\end{eqnarray}
where $m_e$ is the electron mass, $q_e$ is the electron charge,  the bulk Lorentz factor $\Gamma$ is introduced by transferring the shock co-moving to the observer rest frame and $B'$ is the magnetic field strength in the shock co-moving frame.  In GRB problems, one usually assumes the magnetic field density ($B'^2/8\pi$) is a fraction of the internal energy of post-shocked medium  with a shock equipartition parameter  $\epsilon_B$. Therefore, the co-moving magnetic field is in form of \citep{sari98}
\begin{eqnarray}
B'=(32 \pi m_{p} \epsilon_{B} n)^{1 / 2} \Gamma c.
\end{eqnarray}

The energy distribution of  shock accelerated electron is usually assumed to be a power law, with $N(\gamma_{e})d\gamma_{e}\propto \gamma_{e}^{-p}d\gamma_{e}$. The minimum Lorentz factor of the electron, $\gamma_{m}$ could be obtained by the conservation laws of particle number and energy. Assuming a constant fraction $\epsilon_{e}$ of the post-shock internal energy goes into the electrons, one has \citep{sari98}
\begin{eqnarray}
\gamma_{m}=\epsilon_{e} \left(\frac{p-2}{p-1} \right) \frac{m_{p}}{m_{e}} (\Gamma-1).
\end{eqnarray}

Another critical electron Lorentz factor is the cooling Lorentz factor $\gamma_{c}$.  When $\gamma_{e}>\gamma_{c}$, electrons would loss most of their energies by synchrotron radiation, otherwise the cooling caused by the radiation can be ignored.
If synchrotron radiation is dominated, the cooling Lorentz factor is given by \citep{sari98}
\begin{eqnarray}
\gamma_{c}=\frac{6 \pi m_{e} c}{\sigma_{T} \Gamma B'^{2} t},
\end{eqnarray}
where $t$ refers to time in the rest frame of observer, $\sigma_{T}$ is Thomson cross-section.

As shown in Eq (\ref{Eq:fre}), electrons with different Lorentz factors $\gamma_{e}$ have different radiation frequencies $\nu (\gamma_e)$. Two characteristic frequencies, $\nu_{m}=\nu(\gamma_m)$ and $\nu_{c}=\nu(\gamma_c)$ would determine the synchrotron spectrum. The evolution of these frequencies with time can be derived from shock dynamics.

For a given dynamical time $t$, if $\gamma_c>\gamma_{m}$, it means only a small fraction of electrons could be cooled,  This is called slow cooling regime. In this case, for an observational frequency $\nu$, the synchrotron spectrum is described a broken power law characterized by $\nu_m$ and $\nu_c$ as follows \citep{sari98}
\begin{eqnarray}
F_{\nu}=\left\{\begin{array}{ll}{(\nu / \nu_{m})^{1 / 3} F_{\nu, \max },} & {\nu<\nu_{m}} \\ {(\nu / \nu_{m})^{-(p-1) / 2} F_{\nu, \max },} & {\nu_{m}<\nu<\nu_{c}} \\ {(\nu_{c} / \nu_{m})^{-(p-1) / 2}(\nu / \nu_{c})^{-p / 2} F_{\nu, \max },} & {\nu_{c}<\nu}\end{array}\right.,
\end{eqnarray}
On the other hand, for $\gamma_{c}<\gamma_{m}$, all the accelerated electrons could be cooled in the dynamical timescale $t$. This is named fast cooling regime. The radiation spectrum of the shock is \citep{sari98}
\begin{eqnarray}
F_{\nu}=\left\{\begin{array}{ll} {(\nu / \nu_{c})^{1 / 3} F_{\nu, \max },} & {\nu <\nu_{c}} \\ {(\nu / \nu_{c})^{-1 / 2} F_{\nu, \max },} & {\nu_{c}<\nu<\nu_{m}} \\ {(\nu_{m} / \nu_{c})^{-1 / 2}(\nu / \nu_{m})^{-p / 2} F_{\nu, \max },} & {\nu_{m}<\nu}\end{array}\right.,
\end{eqnarray}
where $F_{\nu,\max}$ is the observed peak flux at luminosity distance $D_L$ from the source, which can be estimated as (\cite{sari98})
\begin{eqnarray}
F_{\nu, \max } = N_{e,\rm{tot}}\frac{m_{e}c^{2}\sigma_{T}}{12\pi q_{e}D_{L}^{2}}B'\Gamma,
\end{eqnarray}
where $N_{e,\rm{tot}}=4\pi nR^3/3$ is total number of swept-up electrons in the post-shock medium.

With the dynamics and radiation equations described above, we can calculate the time-dependent spectra of the blast wave emission and the light curves for any observational frequencies. For the purpose of this work, here we focus on the X-ray light curve within the \textit{Swift}-XRT energy band ($0.3-10$ keV). Note that the dominated radiation mechanism for X-ray afterglow emission is synchrotron radiation {\citep{meszaros97,sari98}, therefore, inverse Compton mechanism is ignored in our calculation.

\subsection{Influence of the Model Parameters on the X-ray Light Curve}

In this subsection, we show the numerical results of our model. Here, we explore the influence of the model parameters on the theoretical X-ray light curves. There are several free parameters in our model, which can be divided into two categories. The first category is related to the BZ process and the newborn BH, including the initial mass of the newborn BH $M_{\bullet,0}$, the initial BH spin $a_0$, the viscosity timescale of disk $\tau_{\rm{vis}}$, peak time of fallback $t_{p}$ and the peak fallback rate $\dot{M}_{p}$. The second category is associated with the external shock, including initial kinetic energy of GRB outflow $E_{0}$, the initial bulk Lorentz factor of GRB outflow $\Gamma_{0}$, the initial opening angle of the jet $\theta_{0}$, the equipartition parameters for the magnetic field and electrons $\epsilon_{B}$ and $\epsilon_{e}$, and the electron distribution index $p$.

For the first category, considering that the newborn BH is produced by the collapse of a supermassive NS, the initial mass of the BH, $M_{\bullet,0}$ should be close to the maximum mass of NS, which should be between 2 to 3 solar mass \citep{lattimer12}. Within this range, the exact value of $M_{\bullet,0}$ hardly affect the BZ power.  Here we adopt $M_{\bullet,0}=2.2M_{\odot}$ as a fiducial value. The influence of $a_0$, $\tau_{\rm{vis}}$ and $t_{p}$ are shown in Figure 2. We find that: 1) a larger value of $a_0$ leads to a more luminous and longer duration plateau; 2) a longer  viscosity timescale  of disk $\tau_{\rm{vis}}$ results in a longer duration, but lower luminous plateau, which is understandable, since according to Eq (\ref{Eq:tau_vis}), a larger value of $\tau_{\rm{vis}}$ corresponds to a slower and weaker BH accretion; 3) a larger value of $t_p$ leads to lower luminosity of lightcurves.  

For the second category, the initial opening angle of the jet hardly affect the final results. Here we adopt $\theta_{0}=0.1$ as a fiducial value. The influence of $\Gamma_{0}$, $n$, $\epsilon_{B}$, $\epsilon_{e}$ and $p$ are shown in Figure 2. We find that: 1) when $n$ is larger, the blast wave collects more material in a higher density ISM, which leads to more luminous lightcurve; 2) since synchrotron  radiation intensity increases with the magnetic energy density and the electrons energy in the external shock, the increase of $\epsilon_{B}$ and $\epsilon_{e}$ would brighten the X-ray flux; 3) the value of $\Gamma_0$ mainly affects the early behavior of the light curve, but not the late behavior when energy injection happens; 4) the value of $p$ mainly affects the decline slope of the light curve.

In order to better reflect the injected energy $E_{\rm inj}$ and the initial kinetic energy of the blast wave $E_0$, we test the influence of parameter $\eta \equiv E_{\rm inj}/E_0$ instead of $\dot{M_{p}}$ and $E_0$. As expected, the larger of $\eta$ value, the more significant of the light curve re-brightening (see Figure 2).

\begin{figure}[!htb]
\centering
\includegraphics[width=5.5cm,height=6cm]{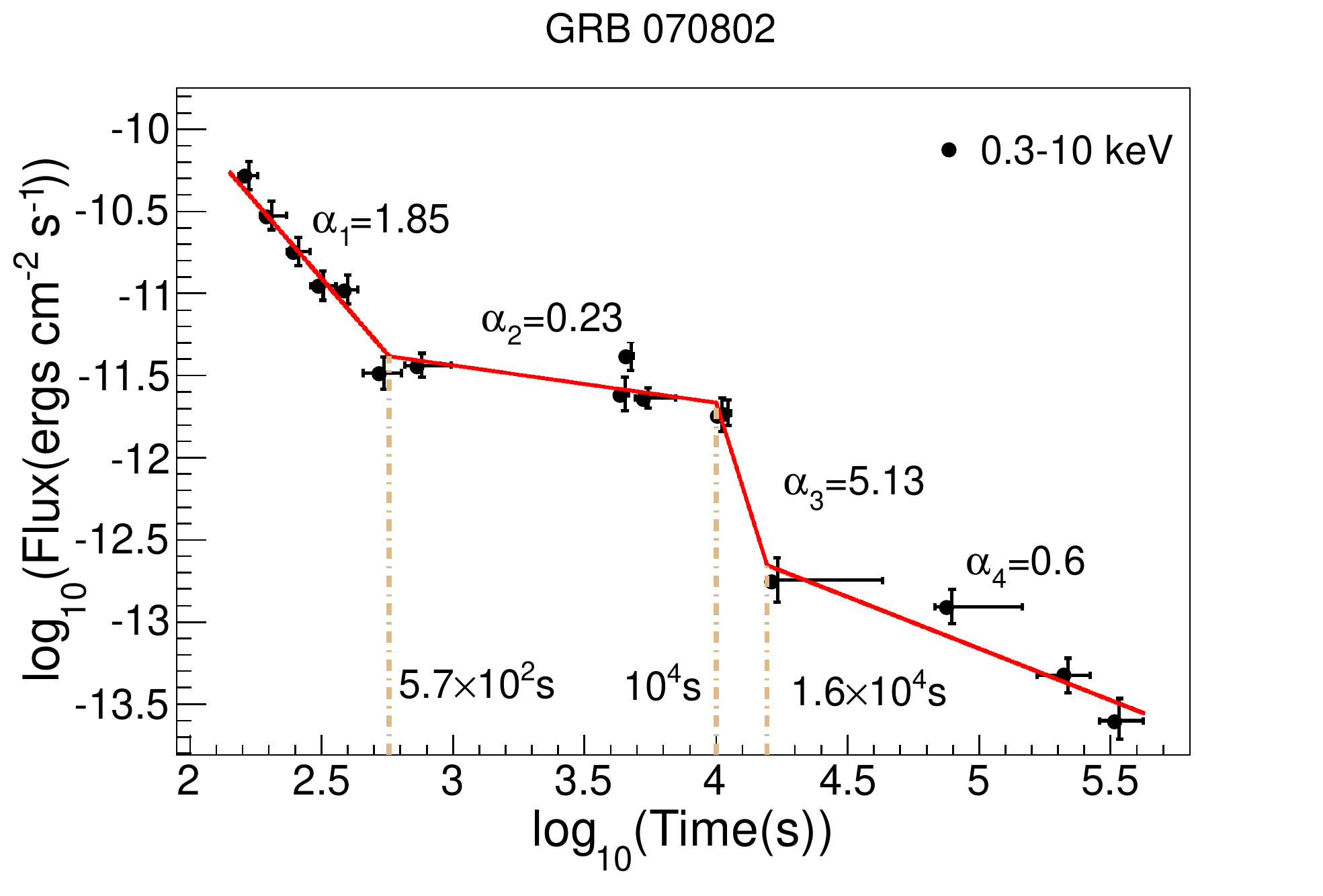}
\includegraphics[width=5.5cm,height=6cm]{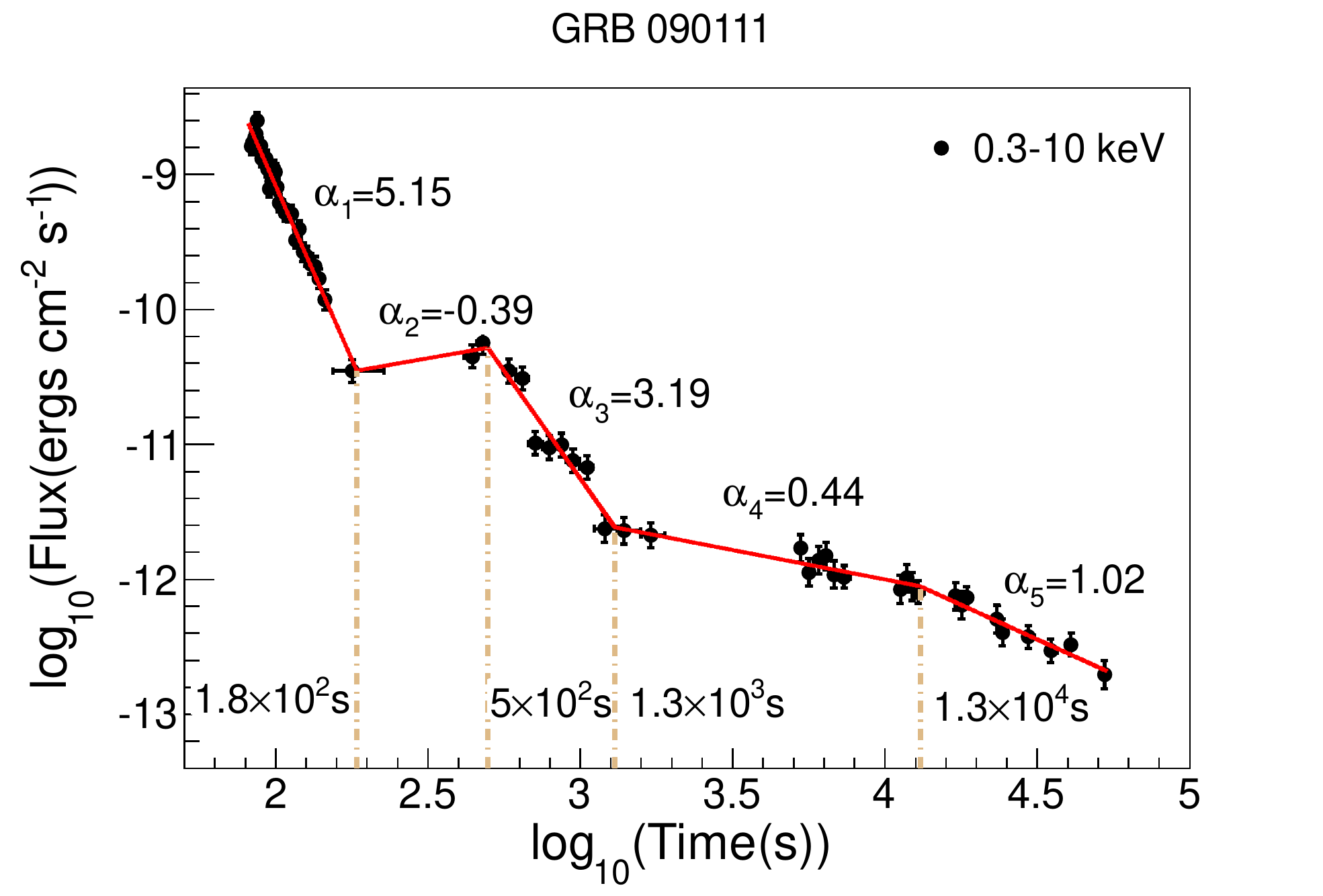}
\includegraphics[width=5.5cm,height=6cm]{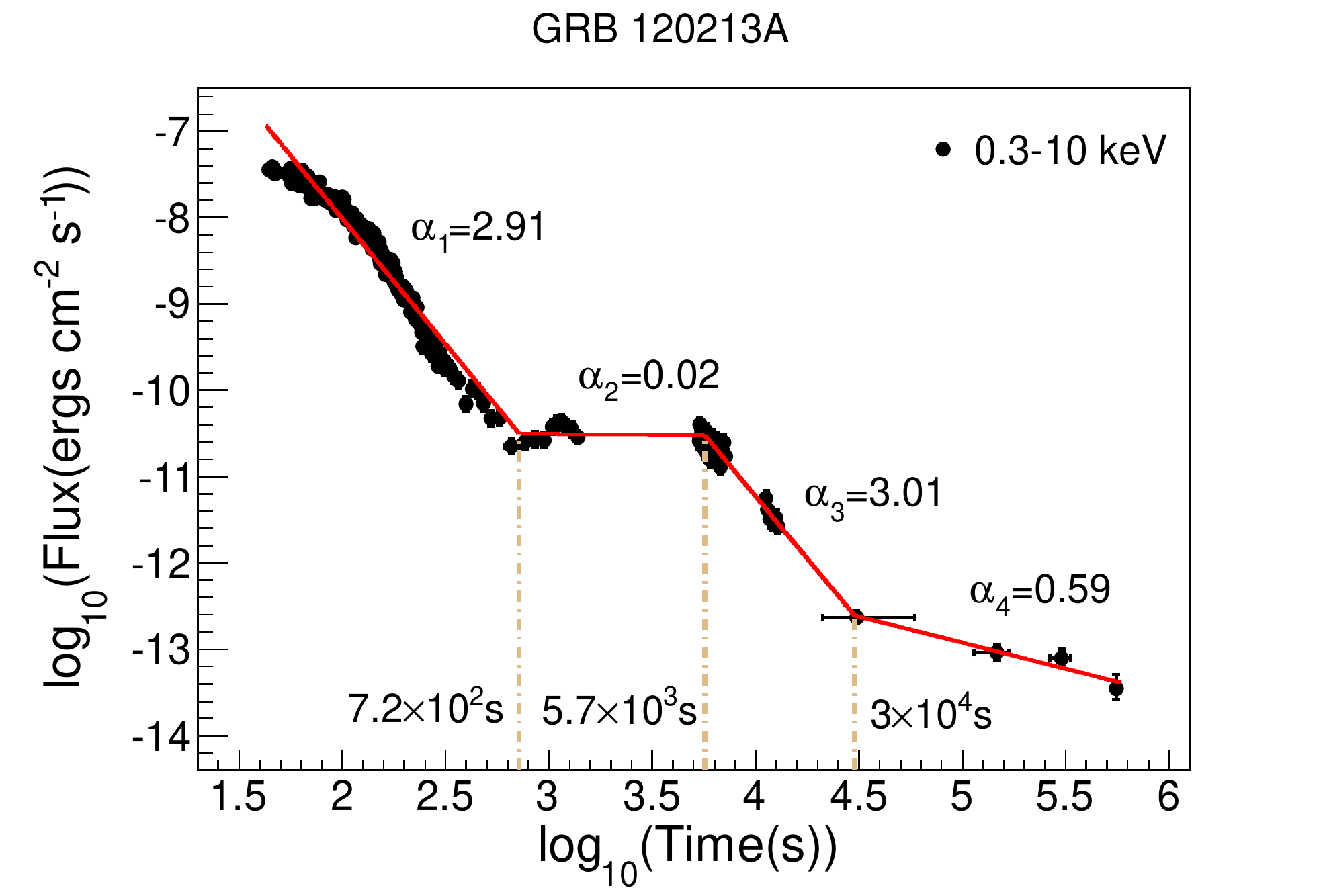}
\caption{ Temporal fitting results of XRT 0.3-10 keV light curve for GRB 070802, 090111 and 120213A by using MARS technique.}
\label{fig-1}
\end{figure}

\section{Sample Selection and Interpretation}
\subsection{Data Reduction And Sample Selection}{\label{data reduction and sample selection}}

For the purpose of this work, we systemically search  sources consisting two X-ray plateaus, where the first one should be an ``internal plateau''  followed by a steep decay.  The XRT light curves data were downloaded from the \emph{Swift}/XRT team website\footnote{\url{http://www.swift.ac.uk/xrt\_ curves/}} \citep{evans07,evans09}, and processed with HEASOFT v6.12.  There were 1291 GRBs  detected by \emph{Swift}/XRT between 2004 February and 2017 July, with 625 GRBs have well-sampled XRT light curves, which including at least 6 data points, excluding the upper limit. XRT light curves for all selected sample are then fitted with multi-segment broken power law function (in logarithmic scale). Here we adopt the multivariate adaptive regression spline (MARS) technique (e.g. \cite{friedman1991}) to fit the light curves. MARS technique can automatically determine both variable selection and functional form, resulting in an explanatory predictive model. Some previous works have proven that MARS can automatically fit the XRT light curve with multi-segment broken power-law function (results in general consistent with fitting results provided by the XRT GRB online catalog \citep{evans07,evans09}), detect and optimize all breaks, and record all break times and power-law indices for each segment (see \cite{zhang14} and \cite{zhao19} for details). Here we treat the adjacent segments with index difference smaller than 0.3 as one component when calculating the segment time span, in order to avoid the potential over-fitting problem from MARS technique. With the fitting results provide by MARS, we searched for candidates having two shallow decay components, where segments with decay slope shallower than 0.65 and time span in log scale larger than 0.4 dex are defined as the shallow decay components \footnote{Based on the early-year Swift observations, \cite{liang07} performed a systematic analysis for the shallow decay component of GRB X-ray afterglow, and they found that the distribution of shallow decay slope ($\alpha_s$) is normal distribution, that is  $\alpha_s=0.35\pm0.35$ (quoted errors are at the 1-$\sigma$ confidence level). Recently, \cite{zhao19} revisited the analysis with an updated sample and they found that with a larger sample, the distribution of the $\alpha_s$ is still normal distribution with $\alpha_s=0.43\pm0.22$. In this work, we adopt 1-$\sigma$ region upper boundary of $\alpha_s$ as the selection criteria for the first and second plateau feature.}. For the purpose of this work, we add one more criteria for the sample selection, i.e., the decay slop following the first shallow decay component should be $\gtrsim$3. Eventually, we find 3 long GRBs (GRB 070802,GRB 090111 and GRB 120213A) meeting all our requirement.

GRB 070802 triggered the BAT at 07:07:25 UT on 2007 August 2. $T_{90}$ is 16.4 s$\pm$ 1.0 s. The time-averaged spectrum from $T+4.9$ s to $T+23.2$ s is best fitted by a single power law (SPL) function and the power law index of spectrum is $\Gamma_{\gamma}=1.79 \pm 0.27$. The fluence in the 15-150 keV band is $S_{\gamma}= 2.8 \pm 0.5 \times 10^{-7} \rm{ergs}/\rm{cm}^{2}$ (\cite{cummings07}).   The XRT  began taking data  138 s after the  trigger \citep{barthelmy07}. The fitting result of XRT data provided by MARS is shown in Figure \ref{fig-1}. It is worth noting that on this case, the fitting curve given by the on-line XRT catalogue is a little different from that given by MARS. Since there are few data points in the late stage, the on-line catalogue take all the data after the first plateau into one segment, which is considered as a normal decay component. According to its own algorithm, however, MARS automatically fits the late data with two segments, and the second segment has a decline slope less than 0.65. In this case, we take GRB 070802 as one of our candidates, but with relatively weak evidence. The X-ray fluence and photon index for the two shallow decay segments are listed in Table \ref{table-1}.} The UV/Optical Telescope (UVOT; \cite{roming05}) start to collect data 141 s after the trigger.  No new source in the UVOT observations at the location of the refined XRT position (\cite{kuin07}). \cite{prochaska07} observed the afterglow of GRB 070802 with the ESO VLT + FORS. From the detection of several Fe lines in a 30 minutes spectrum starting on August 2.378 UT, the redshift was measured as $z=2.45$. {\cite{kruhler08} presented the optical and near-infrared photometry of the afterglow obtained with the multichannel imager GROND. Unfortunately, the late optical data points are also scarce, and there is no data point around the time when the second X-ray plateau emerges. The late optical-IR data could be basically consistent with a single decay segment, so that the optical emission trend might be different with the X-ray band, just like what is found for the case of GRB 070110 \citep{troja07}. On the other hand, as shown in the next section, the late optical-IR data could also be well fitted by our proposed model simultaneously with the X-ray data.

The BAT triggered and located GRB 090111 at 23:58:21 UT on 2009 January 11.  $T_{90}$ is 24.8$\pm$ 2.7 s. The time-averaged spectrum from $T-2.9$ s to $T+25.6$ s is best fitted by a SPL function.  The power law index of the time-averaged spectrum is $\Gamma_{\gamma}=2.37 \pm0.17$.  The fluence in the 15-150 keV band is $S_{\gamma}=6.2\pm0.6\times 10^{-7}\ \rm{ergs}/\rm{cm}^{2}$ (\cite{Stamatikos09}). XRT observations started at 76.6s after the trigger \citep{Hoversten09}. The fitting result of XRT data provided by MARS is shown in Figure \ref{fig-1}. The X-ray fluence and photon index for the two shallow decay segments are listed in Table \ref{table-1}. Note that GRB090111 has data more difficult to interpret mainly due to the orbital gap around thousands of seconds, where the data could also be interpreted as a flare followed by a decay (this would also be consistent with the variation in hardness ratio seen in the XRT data).The UVOT to collect data starting 86 s after the trigger. No source was detected by the UVOT at the X-ray afterglow position (\cite{Hoversten09}). No prompt ground-based observation was reported, probably due to the vicinity ($46^{\circ}$) to the Sun (\cite{margutti09}).

GRB 120213A triggered the BAT at 00:27:19 UT on 2012 February 12.  $T_{90}$ is 48.9$\pm$ 12 s. The time-averaged spectrum from $T-6.31$ s to $T+74.46$ s is best fitted by a SPL function.  The power law index of the time-averaged spectrum is $\Gamma_{\gamma}=2.37\pm0.09$.  The fluence in the 15-150 keV band is $S_{\gamma}=1.9\ \pm  0.1 \times 10^{-6}\ \rm{ergs}/\rm{cm}^{2}$ (\cite{baumgartner12}). The XRT began collect data 54 s  after the trigger \citep{oates12}. The fitting result of XRT data provided by MARS is shown in Figure \ref{fig-1}. The X-ray fluence and photon index for the two shallow decay segments are listed in Table \ref{table-1}. For this case, there is a fairly clear evidence for a second shallow decay component, but unfortunately the data do not exist to know when the shallow decay stops and at what slop the even later time emission decays. The UVOT to collect data starting  58 s after the trigger. No optical afterglow consistent with the XRT position and no prompt ground-based observation was reported (\cite{oates12}).

\begin{deluxetable}{ccCcccl}[b!]
	\tablecaption{The X-ray fluence of the plateau and photon index of the plateau and its follow-up segment}\label{table-1}
	\tablecolumns{8}
	\tablenum{1}
	\tablewidth{0pt}
	\tablehead{
		\colhead{name} &
		\colhead{$S_{X}$\tablenotemark{a}}&
		\colhead{$\Gamma_{X,1}$\tablenotemark{a}}&
		\colhead{$\Gamma_{X,2}$\tablenotemark{a}}&\colhead{$S_{X}$\tablenotemark{b}}&
		\colhead{$\Gamma_{X,1}$\tablenotemark{b}}&\colhead{$\Gamma_{X,2}$\tablenotemark{b}}\\
		\colhead{} & \colhead{$10^{-7}\rm{ergs} \ \rm{cm}^{-2}$}& \colhead{}& \colhead{}& \colhead{$10^{-7}\rm{ergs} \ \rm{cm}^{-2}$}& \colhead{}
	}
	\startdata
	GRB 070802&0.25$\pm$0.1&$2.4^{+0.76}_{-0.63}$&$1.81^{+0.85}_{-0.55}$&0.23$\pm$0.4&$1.81^{+0.85}_{-0.55}$&... \\
	GRB 090111&0.03$\pm$0.2&$1.79^{+1.36}_{-0.69}$&$3.56^{+0.54}_{-0.48}$&0.15$\pm$0.02&$2.19^{+0.53}_{-0.48}$&$2.14^{+0.37}_{-0.33}$\\
	GRB 120213A&1.5$\pm$1&$1.66^{+0.26}_{-0.20}$&$2.34^{+0.2}_{-0.19}$& 0.42$\pm$0.3&$1.74^{+0.69}_{-0.34}$&...\\
	\enddata
	\tablenotetext{a}{ For the first plateau.}
	\tablenotetext{b}{ For the second plateau.}
\end{deluxetable}

\subsection{Model Application to Selected GRBs}{\label{mcmcsimulation}}

In this section, we apply the model described in Section 2 to interpret the X-ray light curve data of GRBs 070802, 090111, and 120213A. Since the first X-ray plateau data could be well explained with magnetar spin-down power \citep{troja07,lyons10,rowlinson10,rowlinson13,luzhang14,lu15}, here we focus on fitting the second X-ray plateau data by consider that the second plateau is produced by energy injection from the new born BH driving BZ power. In order to minimize the $\chi^2$ of the fitting, a Markov Chain Monte Carlo (MCMC) method is adopted. In our MCMC fitting, \textrm{emcee} code \citep{Formanmackey13} is used with a walkers number 160 and $10^4$ burn-in iterations in the ensemble. Considering that the total number of observational data points available are not enough to constrain all the model parameters. In order to reduce the number of the free parameters in our fitting, we fix several parameters at their typical values. For instance, we set $E_{0}=10^{52}\ \rm{ergs}$,  $\Gamma_{0}=100$, $n =1\ \rm{cm}^{-3}$, $\theta_{0}=0.1$,$\epsilon_{e}=0.1$, $\epsilon_{B}=10^{-4}$, $p=2.5$, and we only take the initial BH spin $a_0$, the viscosity timescale of disk $\tau_{\rm vis}$, the peak time of the fallback $t_p$, and the ratio between the injected energy and initial kinetic energy $\eta$ as free parameters. We set the allowed ranges for the four free parameters in our fitting as: $a_0 \equiv[0,1]$, $\log_{10} \tau_{\rm vis} \equiv[t_0,t_1]$, $\log_{10} t_{p} \equiv[t_0,t_1]$, $\log_{10} \eta \equiv[-3,3]$. Here, $t_{0}$ is the ending timescale of internal plateau, which can also be used as the start time of the fall-back accretion and $t_1$ is the time of last observational data point, the values of them for each burst are shown in Figure \ref{fig-1}. 

Figure 4 shows our fitting results for 3 selected GRBs, where the upper panel shows the fitting light curves and the lower panel shows the corresponding corner plot of the posterior probability distribution for the fitting. We can see that, the second plateau for all 3 GRBs could be well fitted with our proposed model. GRB 070802 has a redshift measurement $z=2.45$, and the fitting results at 1$\sigma$ confidence level are are $a_{0}=0.68^{+0.15}_{-0.24}$,  $\log_{10} \tau_{\rm{vis}}= 4.50^{+0.44}_{-0.36}$s , $\log_{10}\eta=1.95^{+0.54}_{-0.38}$ and $\log_{10} t_{p}=4.12^{+0.14}_{-0.08}$ s. It is interesting to note that the late optical-IR afterglow data of GRB 070802 could also be well fitted by our proposed model simultaneously with the X-ray data (in Figure 4, we show the data and fitting result for $Ks$ band as an example). For GRB 090111 and GRB 120213A, due to the lack of redshift measurement, here we adopt $z = 1$ in our analysis. In this case, for GRB 090111, the model parameters at 1$\sigma$ confidence level are  $a_{0}=0.7^{+0.14}_{-0.18}$,  $\log_{10} \tau_{\rm{vis}}= 4.09^{+0.08}_{-0.08}$, $\log_{10}\eta=1.10^{+0.48}_{-0.39}$ and $\log_{10} t_{p}=2.79^{+0.19}_{-0.10}$; for GRB 120213A, the results are $a_{0}=0.69^{+0.14}_{-0.19}$,  $\log_{10} \tau_{\rm{vis}}= 5.43^{+0.18}_{-0.23}$, $\log_{10}\eta=1.63^{+0.44}_{-0.36}$ and $\log_{10} t_{p}=3.95^{+0.26}_{-0.17}$.

From the fitting results, we find that the constraints on model parameters are relatively loose, mainly due to the lack of enough high quality observation data. Even in this case, some general conclusions could still be made: for all 3 GRBs, 1) the fallback accretion model could easily explain the second X-ray plateau data with fairly loose parameter requirements. Note that in the fitting, we have fixed several model parameters, which means the parameter constraints could become even looser if these parameters were also released. 2) The constraints on $\eta$ are relatively tight. $\eta$ was constrained to the order of 10-100, inferring that the BZ power much be 10 or 100 times larger than the initial GRB blast wave kinetic energy, in order to produce the second X-ray plateau feature. 3) Although the allowed parameter space are wide, $a_0$ tends to have a large value, the distribution peaks are larger than 0.6, which is expected since larger $a_0$ would easily give larger BZ power. 4) The distributions of the viscosity timescale of disk $\tau_{\rm vis}$ are also wide. The distribution peaks of $\tau_{\rm vis}$ are relatively large, inferring that the fall-back accretion all falls into the slow accretion regime. 5) The distributions for the peak time of the fallback $t_p$ tend to peak around the ending time of internal plateau, which are around $10^3-10^4$ s. Taken this as the start time of the fall-back accretion, the minimum radius around which matter starts to fall back could be estimated as $r_{\rm fb} \sim 2.8 \times 10^{11}(M_{\bullet}/2.2M_{\odot})^{1/3} (t_0/ \rm{10^4 s})^{2/3}$, which is consistent with the typical radius of a Wolf-Rayet star. 

The fitting mass fallback rate $\dot{M}_\mathrm{fb}$ for these three GRBs reaches the peak value around $10^{-4} M_\odot\ s^{-1}$ at the time about $10^3$ - $10^4\ s$. It is interesting to check whether such a mass fallback rate at that time could be supplied by the progenitor envelope. Here we estimate the mass supply rate from the envelope with the presupernova structure models (e.g., \citealt{Suwa11}; \citealt{WH12}; \citealt{Matsu15}; \citealt{Liu2018}), i.e., 
\begin{equation}\label{key}
\dot{M}_\mathrm{pro} = \frac{dM_r}{dt_\mathrm{ff}} = \frac{dM_r/dr}{dt_\mathrm{ff}/dr} =  \frac{2M_r}{t_\mathrm{ff}}\left(\frac{\rho}{\bar{\rho}-\rho}\right),
\end{equation}
in which $\bar{\rho} = 3M_r/(4\pi r^3)$ is the average density within radius $r$, $M_r$ is the mass coordinate of a shell, $t_\mathrm{ff} = \sqrt{3\pi/(32G\bar{\rho})}=\sqrt{\pi^2 r^3/(8G M_r)}$ denotes the free-fall timescale. By taking some representative progenitor density profiles with different metallicities and masses from \cite{Liu2018} and the references therein, we reproduce the mass supply rate changing with time for those progenitor models. In calculation, we set the time when the central accumulated mass reaches $M_r = M_{0} = 2.2M_\odot$ (our fiducial value of the mass of a new born magnetar) as the zero time reference point, i.e., we take $t = t_\mathrm{ff}(r) - t_\mathrm{ff}(r_0)$, $M_r = M_{0} + \int_{r_0}^{r} 4\pi r^2 \rho dr$, here $r_0$ is the radial coordinate where the enclosed mass is $M_{0}$. As shown in Figure \ref{fig:mdot}, we find that our fitting resulted mass fallback rates are compatible with the theoretical mass supply rate of some low metallicity massive progenitor stars such as those ones with ($Z\lesssim10^{-1}$, $M\gtrsim80M_\odot$), ($Z\lesssim10^{-2}$, $M\gtrsim40M_\odot$), ($Z\lesssim10^{-4}$, $M\gtrsim20M_\odot$), and so on. While the solor metallicity stars might not be so possible to play as the progenitors for our sample.

Based on the fitting results, the magnetic field strength of the new-born BH ($B_\bullet$) for these three GRBs reaches the peak value around $10^{13-14} $G at the time about $10^3$ - $10^4\ s$. According to the dipole spin-down model, one can make estimation for the surface magnetic field $B_{\rm p}$ and the initial spin period $P_0$ of the rapidly spinning magnetar with the first plateau data for all 3 GRBs in our sample \citep{rowlinson13,lu15}. Here we adopt the constant values of the moment of inertia $I=1.5\times10^{45}~{\rm g~cm^{2}}$ and radius R=10 km for a typical neutron star. For GRB 070802, the plateau luminosity and break time are $L_{b}=1.4\times10^{47}~{\rm erg~s^{-1}}$, $t_b=1\times10^{4}$ s, respectively, one can thus derive\footnote{Here we take the break time $t_b$ of the first plateau as the lower limit of characteristic spin-down time, and take the plateau luminosity $L_{b}$ as the characteristic spin-down luminosity \cite[see detailed methods in][]{lu15}. Redshift for GRB 070802 is taken as z = 2.45 and the redshift for GRB 090111 and GRB 120213A is taken as 1.} $P_{0}<4.64\times 10^{-3}~\rm s$ and $B_{\rm p}<2.59\times 10^{15}~\rm G$. For GRB 090111, the plateau luminosity and break time are $L_{b}=2.9\times10^{47}~{\rm erg~s^{-1}}$,  $t_b=427$ s, respectively, one can derive $P_{0}<1.55\times 10^{-2}~\rm s$ and $B_{\rm p}<4.19\times 10^{16}~\rm G$. For GRB 120213A, the plateau luminosity and break time are $L_{b}=2.2\times10^{47}~{\rm erg~s^{-1}}$, $t_b=4.57\times10^{3}$ s, respectively, one can derive $P_{0}<5.44\times 10^{-3}~\rm s$ and $B_{\rm p}<4.49\times 10^{15}~\rm G$. We find that the magnetic field strength of the new-born BH required to power BZ jets is comparable or slightly lower than the magnetic field strength of the magnetar, which is understandable if we consider that the magnetic flux should be roughly conserved when the magnetar collapse into the BH, and some magnetic energy might dissipate due to the interaction between the magnetosphere and the fall back flow \citep{Lloyd19}.

\begin{figure*}
\begin{center}
\setlength{\abovecaptionskip}{0.cm}
\setlength{\belowcaptionskip}{-0.cm}
\hspace{0cm}
\includegraphics[width=5.5cm,height=6cm]{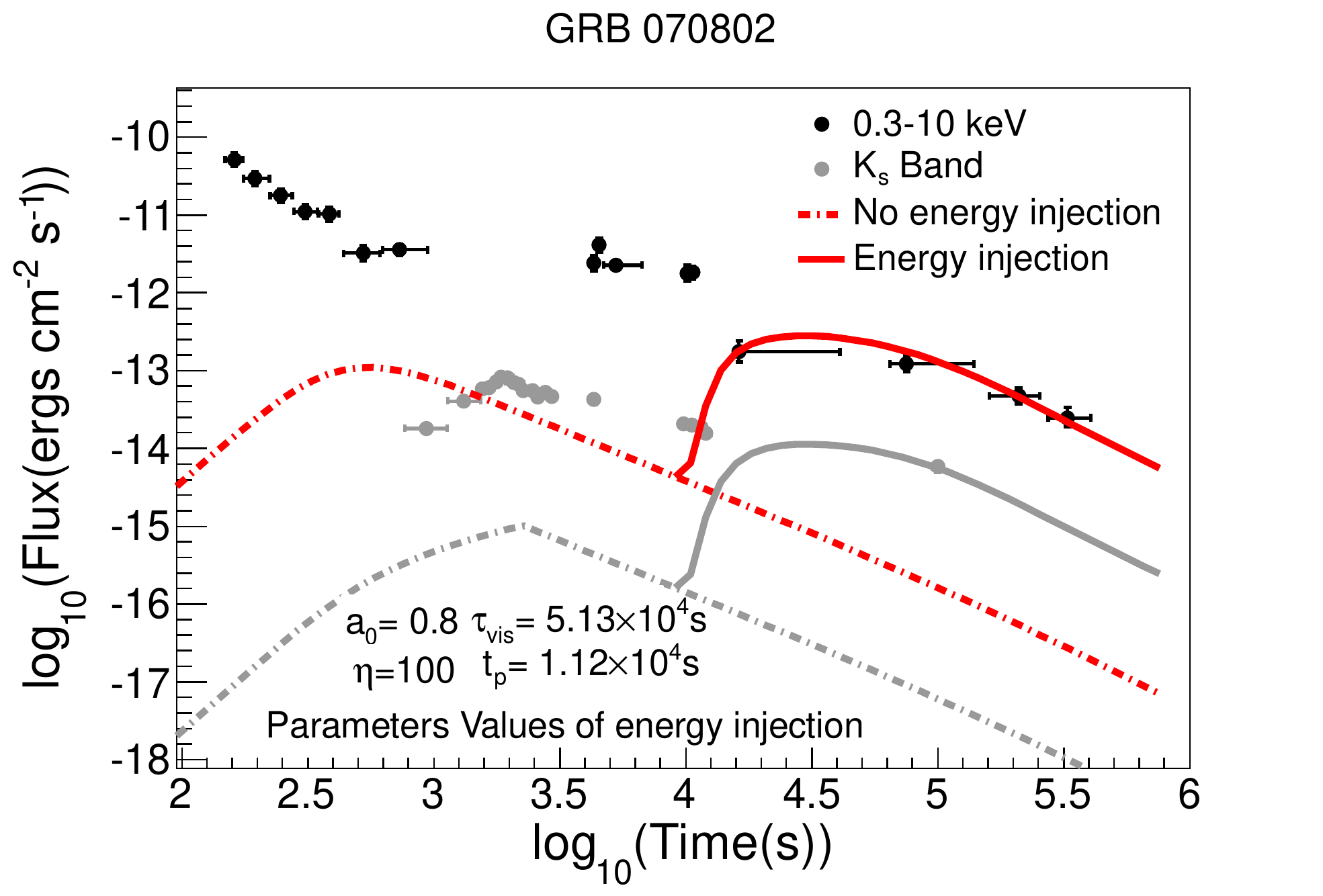}
\includegraphics[width=5.5cm,height=6cm]{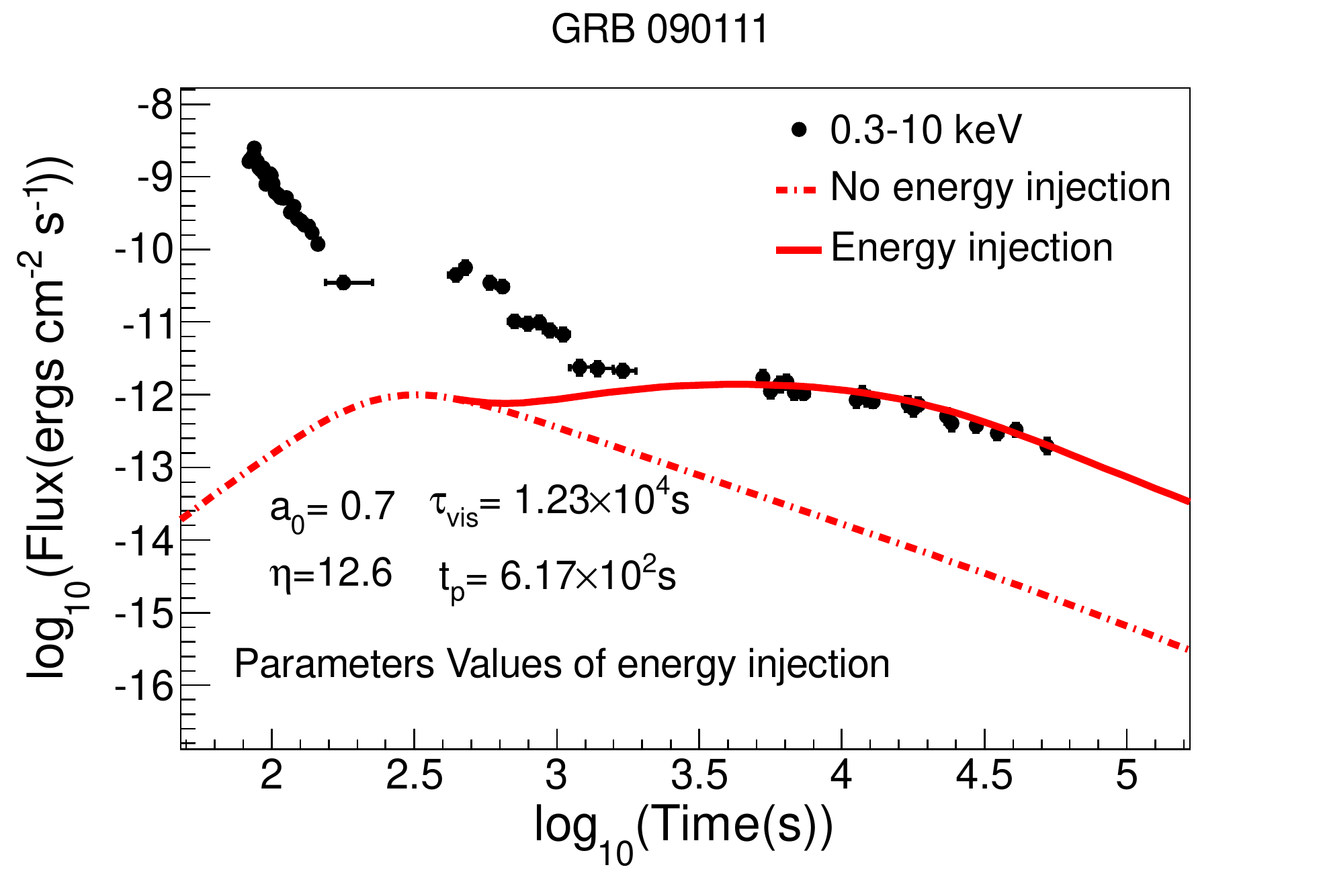}
\includegraphics[width=5.5cm,height=6cm]{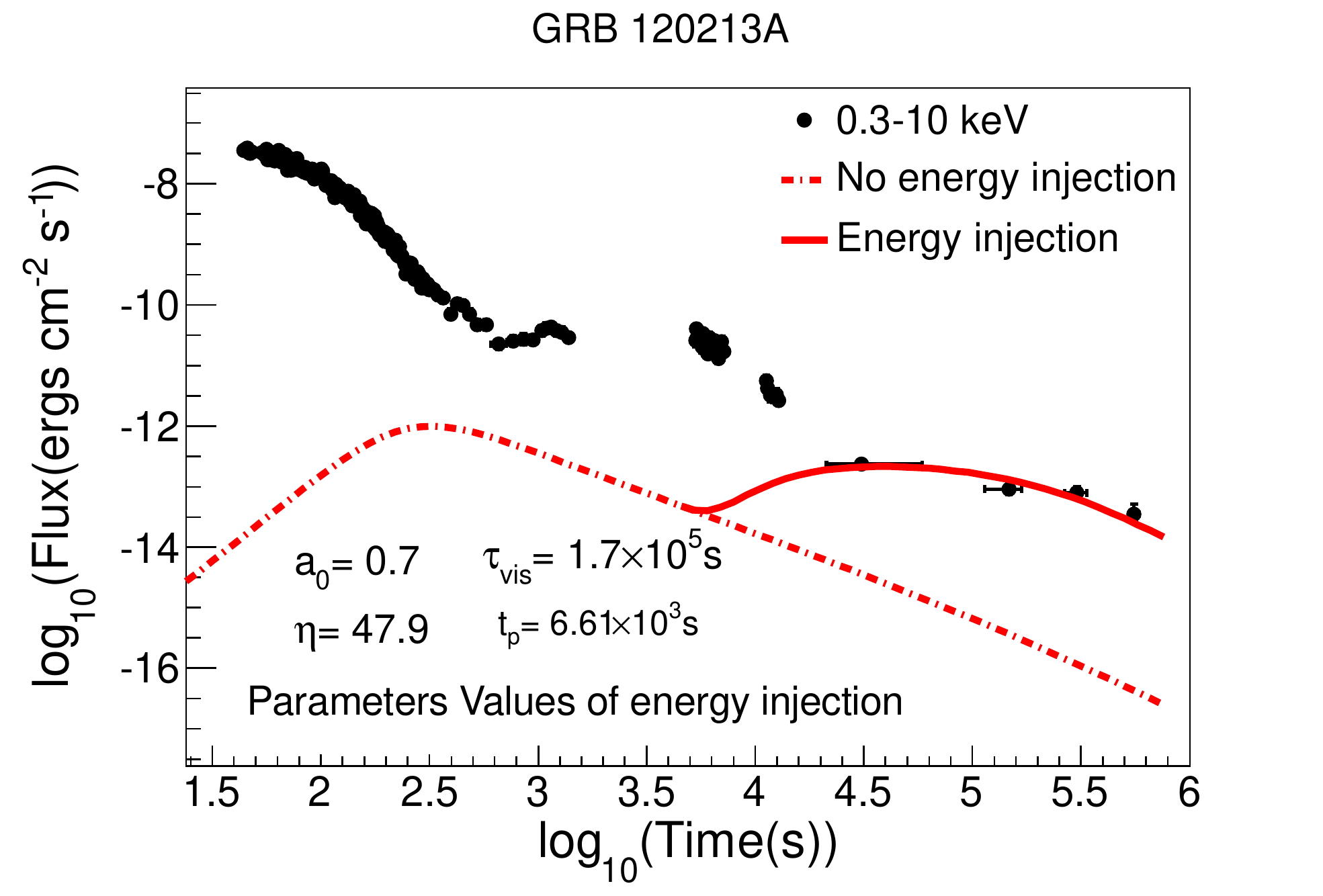}
\includegraphics[width=0.31\textwidth,angle=0]{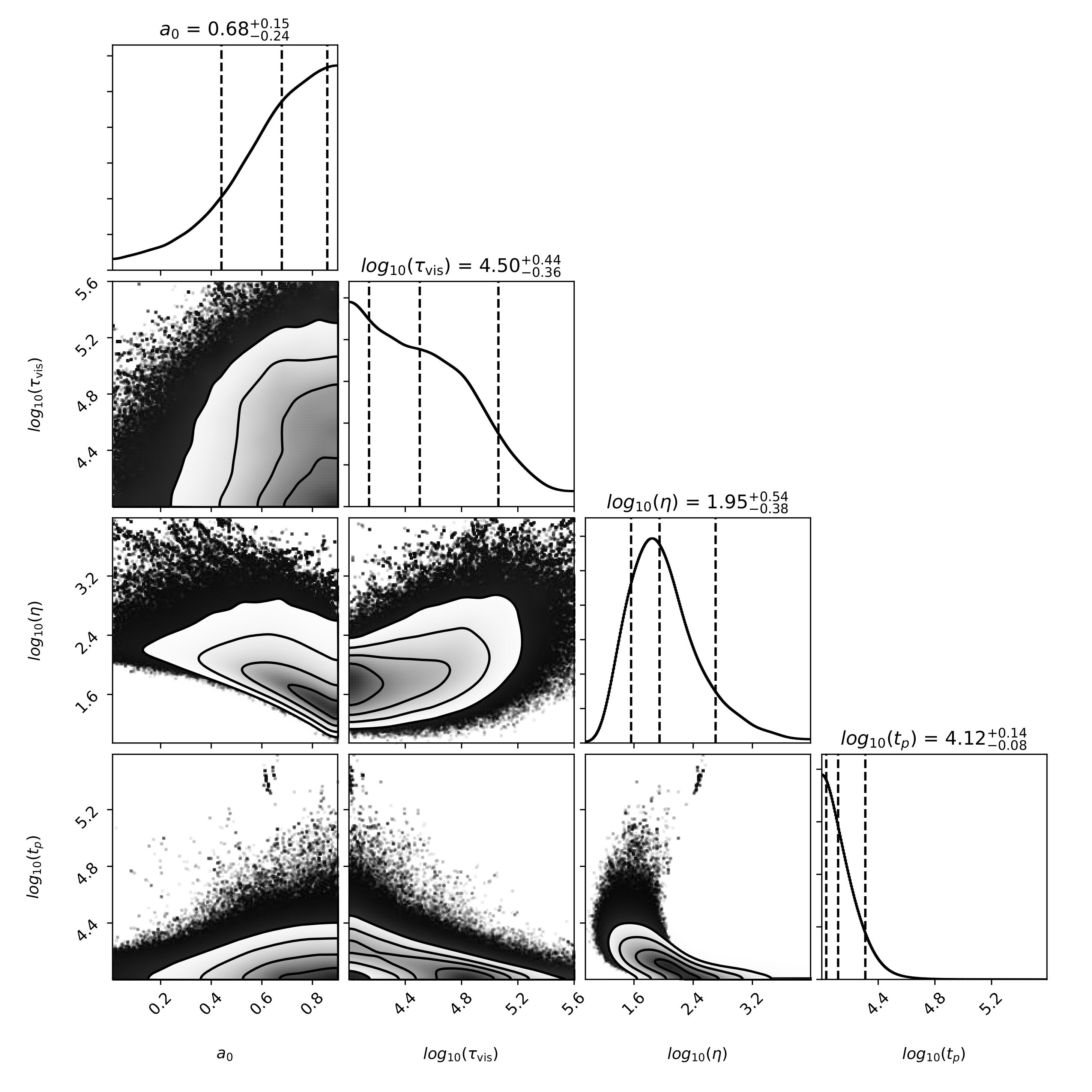}
\includegraphics[width=0.31\textwidth,angle=0]{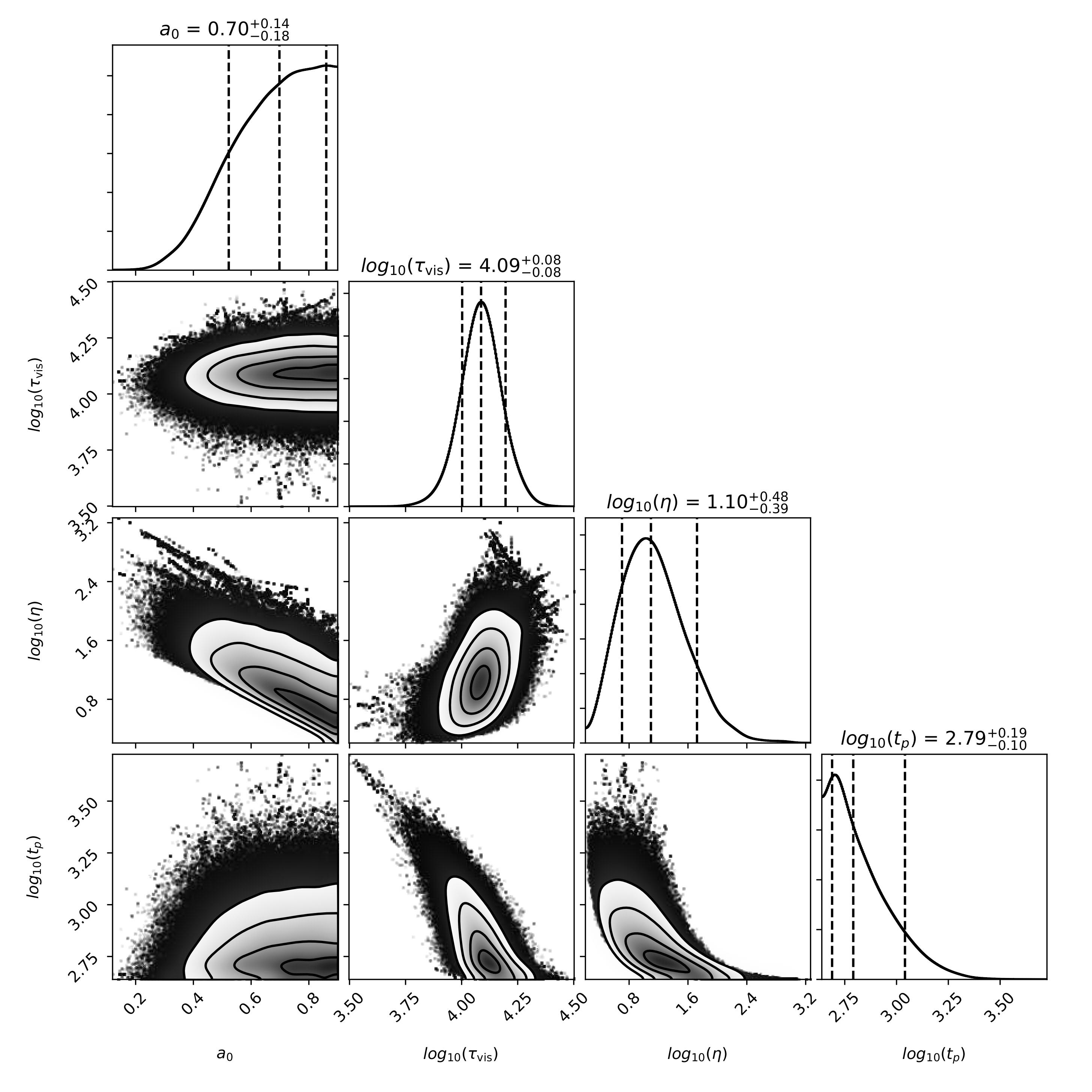}
\includegraphics[width=0.31\textwidth,angle=0]{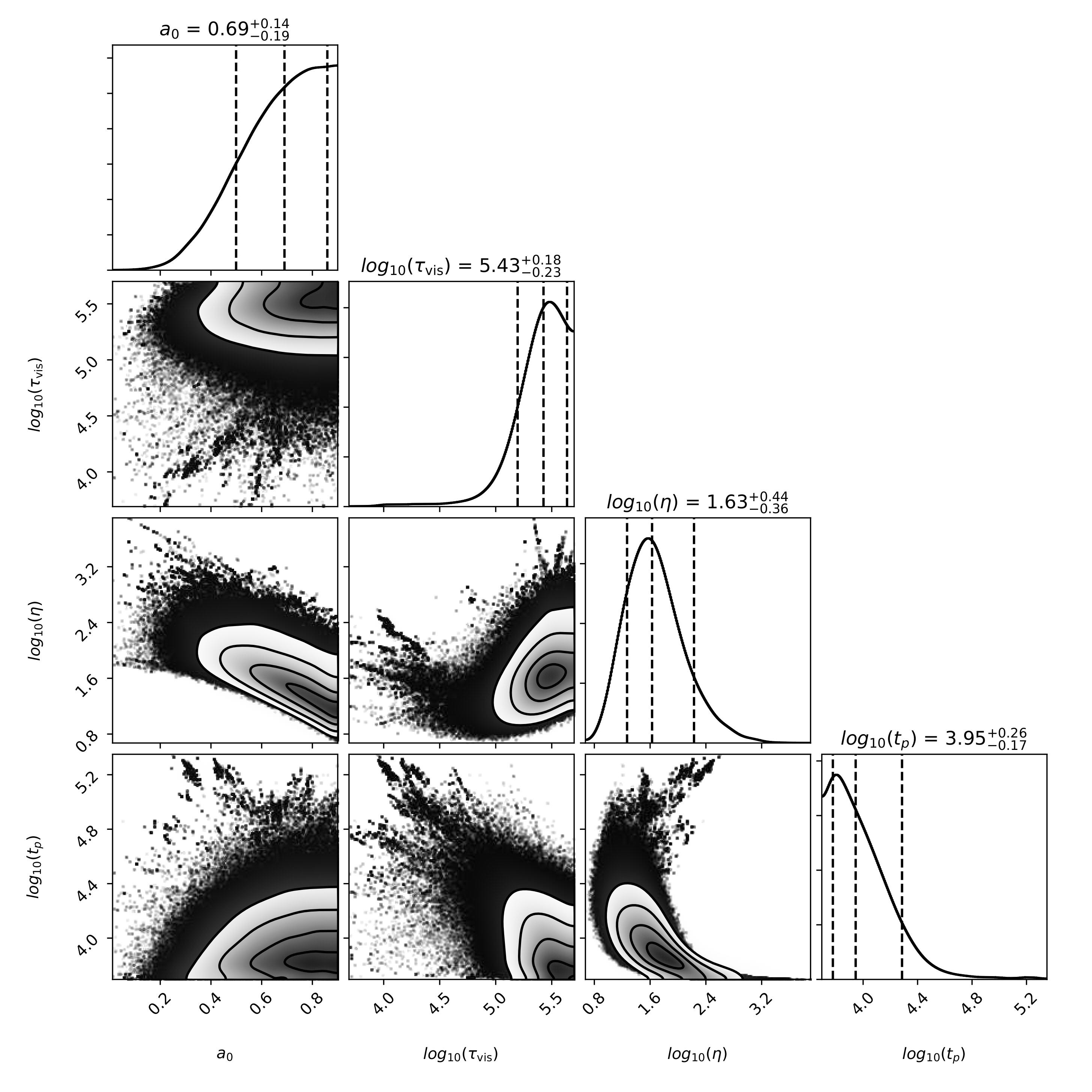}
\caption{Fitting results of the second plateau for GRB 070802, 090111 and 120213A.  }
\label{Fig:MF-LC}
\end{center}
\end{figure*}

\begin{figure}[hbt]
	\centering
	\includegraphics[width=0.5\linewidth]{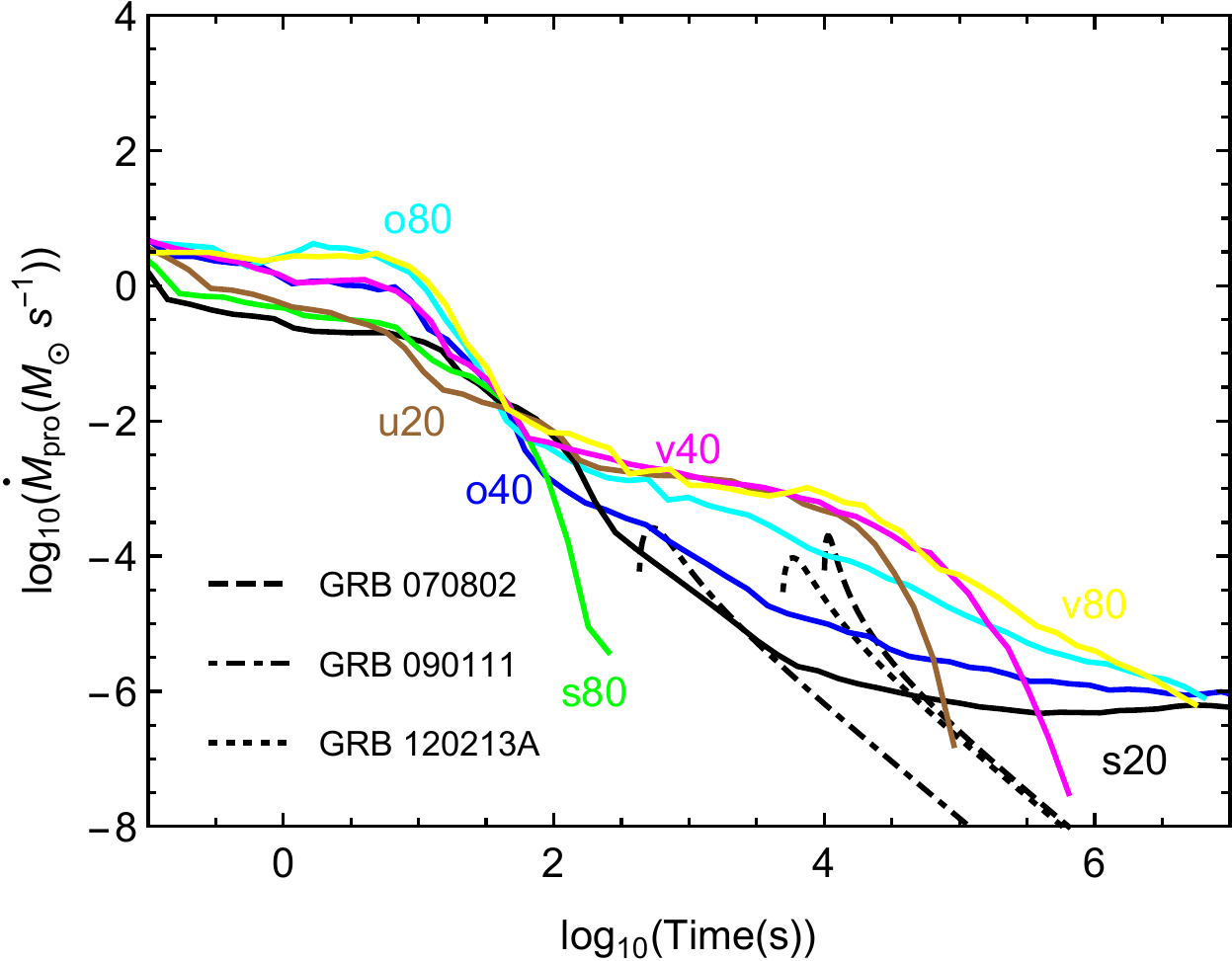}
	\caption{Comparing the fitting mass fallback rate with the mass supply rate of the progenitor stars with different metallicities and masses. The fitting mass fallback rate of the three GRBs, i.e., GRB 070802, GRB 090111, and GRB 120213A are denoted by the dashed line, the dash-dotted line and the dotted line separately. The mass supply rate of the progenitor stars are denoted by the solid colored lines, in which the nomenclature is the same as \cite{Liu2018}, i.e, the signs s, o, v, and u represent the metallicity values $Z = Z_\odot$, $10^{-1}Z_\odot$, $10^{-2}Z_\odot$, and $10^{-4}Z_\odot$, and the numbers beside the signs denote the progenitor masses in unit of solar mass.}
	\label{fig:mdot}
\end{figure}

\section{Discussion and Conclusion}\label{conclusion}

In the 15 years of \emph{Swift}'s operation, it has brought us a lot of observations of GRB X-ray afterglow, which provides valuable information for understanding the GRB central engine. One particular example is the discovery of internal X-ray plateau, (a plateau followed by a very steep decay phase), which is commonly taken as the smoking gun evidence of a rapidly spinning magnetar as the central engine. The very steep decay at the end of the plateau suggests a sudden cessation of the central engine, which is explained as the collapse of a supra-massive magnetar into a black hole when it spins down. If this interpretation is correct, the fall-back accretion from the envelope of progenitor star into the newborn BH could generate some detectable signatures.

Here we propose that the energy extracted from the newborn BH could be continuously injected into the GRB blast wave. In this scenario, We find that with appropriate parameters for the fall back accretion and new born BH, it is possible to produce a second plateau following the steep decay phase of the internal plateau. With a systematical search through the \emph{Swift} -XRT sample, we find three interesting long GRBs, i.e., GRBs 070802, 090111, and 120213A, whose X-ray afterglow light curves contain internal X-ray plateau followed by a second plateau. Here we focus on fitting the second X-ray plateau for these 3 GRBs with our proposed model. We find that in a fairly loose and reasonable parameter space, the second X-ray plateau data could be well interpreted with our model. 

It is worth noting that the quality of current observation data is not good enough to completely eliminate the degeneracy of parameters. In our sample, GRB 090111 has the most observational data points in the period of the second plateau. GRBs 070802 and 120213A only have four data points in the period of the second plateau. Even in this case, some general conclusions could still be reached, for instance, in order to interpret the second plateaus, the initial spin of the new-born BH tends to have a large value (the peak of its posterior probability distribution is larger than 0.6), the later injected energy should be 10 or 100 times larger than the initial GRB blast wave kinetic energy, and the viscosity timescale of disk tends to be large, inferring that the fall-back accretion all falls into the slow accretion regime \citep{lei17}. The mass fall-back rate reaches the peak value around $10^{-4} M_\odot\ s^{-1}$ at the time about $10^3$ - $10^4\ s$, which is compatible with the late mass supply rate of some low metallicity massive progenitor stars. Based on the fitting results, one can infer the total accreted masses $M_{\rm acc} \sim 0.08-0.26M_{\odot}$ and the fallback radii $r_{\rm fb} \sim \rm a~few \times 10^{11}$cm, which is consistent with the typical radius of a Wolf-Rayet star. Combining the X-ray data of the internal plateaus of the three GRBs with the magnetic dipole radiation model of the magnetar, we coarsely estimate the strength of magnetic field of the magnetar before its collapsing to a BH. We find the magnetic field is strong enough to drive a BZ jet.

If our interpretation is correct, the 3 GRBs with two X-ray plateaus provide additional evidence for rapidly spinning magnetar as the GRB central engine. It is worth noticing that most GRBs with “internal plateaus” do not show a second X-ray plateau, which means for most cases the fallback accretion may be relatively weak, so that the injected energy is smaller than initial kinetic energy of GRB blast wave. Future observations are likely to discover more similar events, which could offer more information of the properties of the magnetar as well as the newborn BH.

In this paper, we ignore the mass into the outflow from disk, which will reduce the accretion rate onto BH horizon  during late central engine activity. Usually, the distribution of accretion rate with disk radius is simply described with a power-law model due to the poor knowledge of the disk outflow. The effects of such outflow are thus highly relied on the uncertain  power-law index parameter. The existence of disk outflow may also be important to comprehend the baryon loading into GRB jet \citep{lei13,lei17} and $^{56}\rm Ni$ synthesis for associated supernovae \citep{song19}. We hope future general-relativistic magnetohydrodynamic  (GRMHD) simulation for a better understanding.

We adopt a simple model to describe the evolution of the fall-back accretion rate. For long GRBs, the envelope of the progenitor star is considered as the mass supply of the fall-back accretion \citep{kumar08a}. The evolution of fall-back accretion rate is thus a good tracker of the structure of the progenitor envelope \citep{Liu2018}. We will explore the time-dependent fall-back accretion rate and the expected afterglow lightcurves from long GRBs with progenitors of different masses, angular velocities and metallicities in future, and constrain the characteristics of stars by comparing the second plateau data with our model.

\acknowledgments
We thank Bing Zhang for helpful discussion and the anonymous referee for the helpful comments that have helped us to improve the presentation of the paper. This work is supported by the National Natural Science Foundation of China (NSFC) under Grant No. 11722324,11690024,11633001,11773010 and U1931203, the Strategic Priority Research Program of the Chinese Academy of Sciences, Grant No. XDB23040100 and the Fundamental Research Funds for the Central Universities. LDL is supported by  the National Postdoctoral Program for Innovative Talents (Grant No. BX20190044), China Postdoctoral Science Foundation (Grant No. 2019M660515)  and ``LiYun'' postdoctoral fellow of Beijing Normal University. 
\software{XSPEC(\cite{arnaud96}), HEAsoft(v6.12;\cite{heasarc14}), root(v5.34;\cite{BrunRademakers97}), emcee(v3.0rc2;\cite{Formanmackey13}),corner(v2.0.1;\cite{Foremanmackey16})    }

\clearpage

\end{document}